\documentclass[onefignum,onetabnum]{siamart171218}

\usepackage{lipsum}
\usepackage{amsfonts}
\usepackage{graphicx}
\usepackage{epstopdf}
\usepackage{algorithmic}
\ifpdf
  \DeclareGraphicsExtensions{.eps,.pdf,.png,.jpg}
\else
  \DeclareGraphicsExtensions{.eps}
\fi

\newsiamremark{remark}{Remark}
\newsiamremark{hypothesis}{Hypothesis}
\crefname{hypothesis}{Hypothesis}{Hypotheses}
\newsiamthm{claim}{Claim}

\headers{M.L. Analysis of Complex Networks in Hyperspherical Space}{Mar\'ia Pereda and Ernesto Estrada}

\title{Machine Learning Analysis of Complex Networks in Hyperspherical Space\thanks{Submitted to the editors DATE.
}}

\author{Mar\'ia Pereda\thanks{RWTH Aachen University, Chair for Computational Social Sciences and Humanities, Germany. Grupo Interdisciplinar de Sistemas Complejos, Departamento de Matem\'aticas, Universidad Carlos III de Madrid. Unidad Mixta Interdisciplinar de Comportamiento y Complejidad Social (UMICCS), Spain. \email{maria.pereda@humtec.rwth-aachen.de}}
\and Ernesto Estrada \thanks{Department of Mathematics and Statistics,
University of Strathclyde, 26 Richmond Street, Glasgow, UK. \email{ernesto.estrada@strath.ac.uk}}}

\usepackage{amsopn}

\makeatletter
\newcommand*{\addFileDependency}[1]{
  \typeout{(#1)}
  \@addtofilelist{#1}
  \IfFileExists{#1}{}{\typeout{No file #1.}}
}
\makeatother

\newcommand*{\myexternaldocument}[1]{%
    \externaldocument{#1}%
    \addFileDependency{#1.tex}%
    \addFileDependency{#1.aux}%
}

\usepackage{color}
\usepackage{xcolor}
\usepackage{array}
\usepackage{float}
\usepackage[]{subfig}
\usepackage{textcomp}
\usepackage{multirow}
\usepackage{graphicx}
\usepackage{amsmath}

\ifpdf
\hypersetup{
  pdftitle={Machine Learning Analysis of Complex Networks in Hyperspherical Space},
  pdfauthor={Mar\'ia Pereda and Ernesto Estrada}
}
\fi

\myexternaldocument{ex_supplement}

\begin{document}

\maketitle

\begin{abstract}
A complex network is a condensed representation of the relational
topological framework of a complex system. A main reason for the existence
of such networks is the transmission of items through the entities
of these complex systems. Here, we consider a communicability function
that accounts for the routes through which items flow on networks.
Such a function induces a natural embedding of a network in a Euclidean
high-dimensional sphere. We use one of the geometric parameters of
this embedding, namely the angle between the position vectors of the
nodes in the hyperspheres, to extract structural information from
networks. Such information is extracted by using machine learning
techniques, such as nonmetric multidimensional scaling and K-means
clustering algorithms. The first allows us to reduce the dimensionality
of the communicability hyperspheres to 3-dimensional ones that
allow network visualization. The second permits to cluster the nodes
of the networks based on their similarities in terms of their capacity
to successfully deliver information through the network. After
testing these approaches in benchmark networks and compare them with
the most used clustering methods in networks we analyze two real-world
examples. In the first, consisting of a citation network, we discover
citation groups that reflect the level of mathematics used in
their publications. In the second, we discover groups of genes that
coparticipate in human diseases, reporting a few genes that coparticipate
in cancer and other diseases. Both examples emphasize the potential
of the current methodology for the discovery of new patterns in relational
data.
\end{abstract}



\section{Introduction}
Complex networks represent a vast category of data systems describing
the topological organization of many complex systems, ranging from
social, technological and ecological to molecular ones \cite{Newman_review,Complex_networks_1,Complex_networks_2,networks_dynamics}.
The representation of this type of data as networks provides information
about the topological, spatial, and functional relations of the data.
In mathematical terms, networks are graphs\textendash simple, directed
and/or weighted\textendash in which nodes represent the entities
of the system and edges represent relations between such entities.
The simplest of all the possible representations of networked data
is by means of simple graphs. In this case only the connectivity between entities is captured by the graph, excluding other structural
factors such as directionality, nature of nodes and strength of relations.
Thus, an important challenge in this modeling scenario is to extract
as much information as possible from this reduced representation of
the data. Thus, the use of data analysis techniques, such as machine
learning \cite{MachineLearningComplexNetworks}, is an important
research area of analysis for networked type of data. 

Machine learning stands at developing computational methods for \textquotedblleft learning\textquotedblright{}
with accumulated experiences, either in a supervised or an unsupervised
way \cite{Machine_Learning_1,Machine_learning_2,Machine_Learning_3}.
In supervised learning the inference of concepts from the data is
performed from a training set \cite{Supervised}. Then, the learning
process constructs a mapping function from this training, which can
then be applied to data not ``seem'' before by the model. These
models correspond either to those of classification or regression.
On the other hand, the main goal of unsupervised learning is to reveal
intrinsic structures that are embedded within the data relationships
\cite{Unsupervised}. In this case, the algorithms are designed to
learn solely guided by the structure of the data provided without
any prior knowledge about the data. The typical unsupervised learning
techniques are: clustering \cite{Clustering_1,Clustering_2,Clustering_3,Clustering_4},
outlier detection \cite{Outlier_1,Outlier_2}, dimensionality reduction
\cite{Dimensionality_reduction_1}, and association \cite{Association}.

An area of unsupervised machine learning on networked systems which
has received a great deal of attention is graph/network clustering
\cite{Graph_clustering,Network_communities_1,Network_communities_1.5,Network_communities_2,Network_community_3}.
In general, the problem consists on the unsupervised detection of
groups of nodes\textendash known as communities in network theory
\cite{Network_communities_1,Network_communities_1.5,Network_communities_2,Network_community_3}\textendash which
share more similarity among them than with nodes outside these clusters.
The main interest in network clustering is due to its numerous applications,
making the problem of graph clustering a data-driven task. The most
frequently used definition of community in networks is the one based
on edge density. For instance, in her 2007 overview of graph clustering
Schaeffer \cite{Graph_clustering} recall that ``\textit{it is generally
agreed upon that a subset of vertices forms a good cluster if the
induced subgraph is dense, but there are relatively few connections
from the included vertices to vertices in the rest of the graph}''.
In his seminal overview of 2010 Fortunato \cite{Network_communities_1.5}
pointed out that ``\textit{communities in graphs are related, explicitly
or implicitly, to the concept of edge density (inside versus outside
the community)}''. He makes clear the difference with data clustering
where ``\textit{communities are sets of points which are \textquotedblleft close\textquotedblright{}
to each other, with respect to a measure of distance or similarity,
defined for each pair of points}''. More recently, Silva and Zhao
\cite{MachineLearningComplexNetworks} it their book tacitly define
a community: ``\textit{as a subgraph whose vertices are densely connected
within itself, but sparsely connected with the remainder of the network}''.
However, the complexity of graphs representing real-world systems
is sufficiently large for not having to restrict our definition of clusters to
those based on edge density only. As a data-driven problem our main
task is to design methods that allow the detection of clusters of
nodes/edges which are structurally similar to each other and that
may contain important functional information about the processes taking
place on real-world systems.

\begin{figure}[]
\begin{centering}
\includegraphics[width=0.6\textwidth]{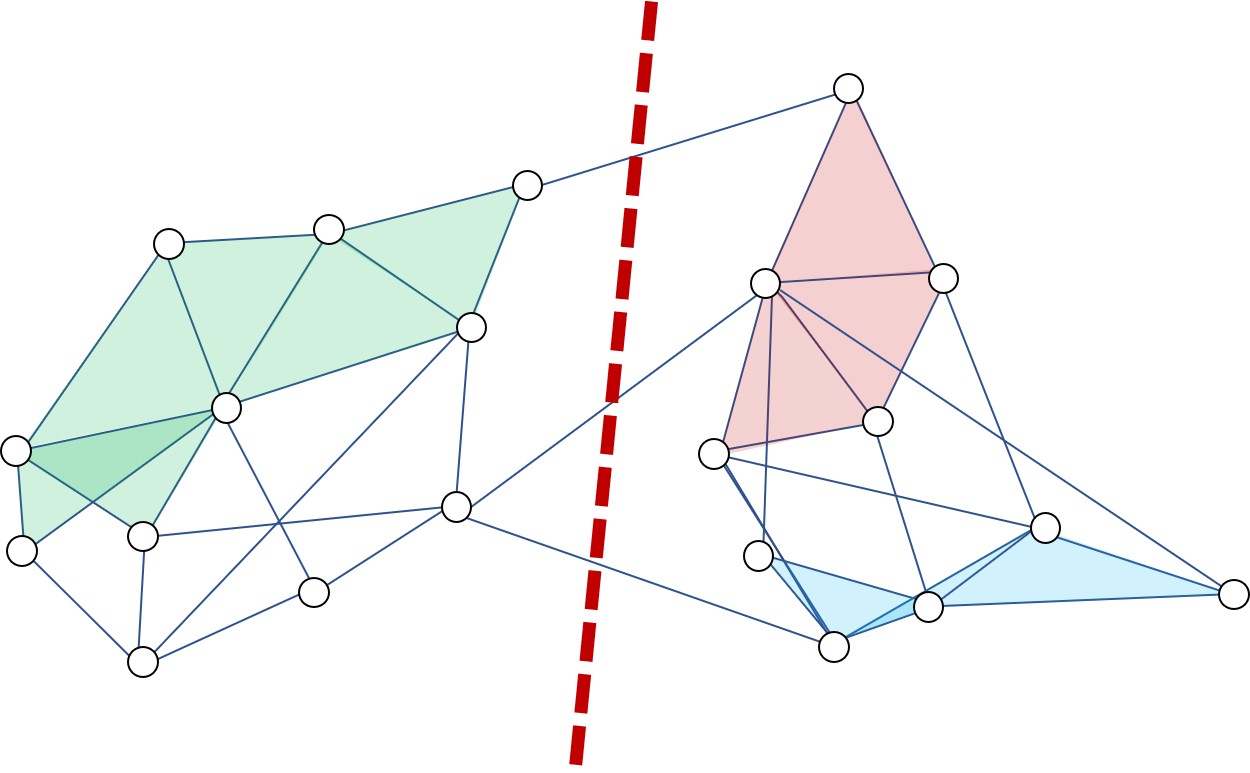}
\par\end{centering}
\caption{Illustration of a hypothetical network in which three clusters based
on fused-triangles exist. Any community detection method based on
cluster density will fail in identifying such clusters as they will
report the existence of two dense communities connected by three edges
only as indicated by the red broken line.}

\label{functional triangles}
\end{figure}

Let us consider here an example for motivating the use of other definitions
of clustering on graphs/networks. Suppose that there is strong empirical
evidence that groups of fused triangles \textendash which can be formally
defined in mathematical terms \textendash represent functional groups
for certain classes of real-world networks. In Fig.\ref{functional triangles}
we illustrate a hypothetical network displaying three clusters of fused
triangles represented in three different colors. Even by eye we can
see that there are two ``communities'' according to the traditional
definition based on edge density. Thus, this means that every method
designed to detect density-communities will fail in detecting the
fused-triangle clusters in this network. It does not mean that a method
designed to detect such triangle-based structures is better or worse
than the ones to detect density-communities. They simply are designed
for performing different tasks on the same dataset.

With the goal of enriching the structural information contained in
graph clustering a series of methods have been proposed which use
the embedding of the graphs in geometric spaces. For instance, Xiao
and Hancock \cite{Xiao_Hancock} embed graphs using the heat-kernel
and then by equating the spectral heat kernel and its Gaussian form
they are able to approximate the Euclidean distance between nodes
on the manifold. After this they perform principal component analysis
(PCA) and demonstrate that it leads to well defined graph clusters.
Other approaches use tools from subspace analysis on a Grassmann manifold
to produce low dimensional representation of the original graphs which
preserves important structural information \cite{Grassman}. Others
embed the networks into hyperbolic space such that network community
structure is obtained from node similarity in such ``\textit{underlying
hidden metric space}'' \cite{Hyperbolic}. In general, these methods
can be grouped under the umbrella of ``geometric learning'' methods
\cite{geometric_learning}. Many of these algorithms are based on
spectral techniques on graphs \cite{Spectral_clustering_1,Spectral_clustering_2}.
Specifically, these approaches propose to embed the vertices of the
original graph into a low dimensional space, which consists of the
top eigenvectors of a special matrix and then carrying out the clustering
in such low dimensional spaces \cite{geometric_learning}.

Here we propose a new approach that can be enclosed in the category of geometric
unsupervised learning. However, instead of ``imposing'' an embedding
of the network in a given manifold we consider the geometric space
generated by the flow of ``items'' on a network in a diffusion-like
process. This space is a Euclidean $(n-1)-$sphere, where $n$ is
the number of nodes of the graph. Using this approach, without the
necessity of dimensionality reduction, we are able to identify clusters
in networks, which not necessarily depend only on their edge density.
After testing the method in a few benchmark networks we embarked in
the analysis of two real-world systems. One is a citation network
and the other a network of gene co-participation in human genetic
diseases. In the first case we discovered the existence of groups
of authors which mainly represent wide-rage of disciplines mainly
demarked by their level of mathematization. In the second example
we discover a few genes which co-participate in neurological diseases
and cancer, as well as in other groups of diseases and cancer. Using nonmetric multidimensional scaling we make a dimensionality reduction of the hyperspheric spaces to 2-spheres, which allow a nice visualization
of the communicability space of networks. 

\section{Preliminaries }

Here we follow standard notation and definitions in network theory
(see for instance \cite{Complex_networks_1}). Let $\Gamma=\left(V,E\right)$
be a simple graph and let $A$ be its adjacency matrix.
We consider here undirected graphs such that the associated adjacency
matrix is symmetric, and its eigenvalues are real. We label the eigenvalues
of $A$ in non-increasing order: $\lambda_{1}\geq\lambda_{2}\geq\ldots\geq\lambda_{n}$.
Since $A$ is a real-valued, symmetric matrix, we can decompose $A$
into 
$A=U\Lambda U^{T}$,
where $\Lambda$ is a diagonal matrix containing the eigenvalues of
$A$ and $U=[\mathbf{\overrightarrow{\psi}}_{1},\ldots,\mathbf{\overrightarrow{\psi}}_{n}]$
is orthogonal, where $\mathbf{\overrightarrow{\psi}}_{i}$ is an eigenvector
associated with $\lambda_{i}$. Because the graphs considered here
are connected, $A$ is irreducible and from the Perron-Frobenius theorem
we can deduce that $\lambda_{1}>\lambda_{2}$ and that the leading
eigenvector $\mathbf{\overrightarrow{\psi}}_{1}$, which will be sometimes
referred to as the \textit{Perron vector}, can be chosen such that
its components $\mathbf{\mathbf{\psi}}_{1}(u)$ are positive for all
$u\in V$. A row of the matrix $U$ corresponding to the node $i$
of the graph is designated here by the vector $\vec{\varphi}_{i}=\left[\mathbf{\mathbf{\psi}}_{1}(i),\ldots\mathbf{\mathbf{\psi}}_{n}(i)\right]^{T}$.

An important quantity for studying communication processes in networks
is the so-called communicability function \cite{Communicability,Estrada_Higham,Estrada_Hatano_Benzi}.
Let $u$ and $v$ be two nodes of $\Gamma$. The communicability function
between these two nodes is defined as

\begin{equation}
G_{uv}=\sum_{k=0}^{\infty}\frac{\left(A^{k}\right)_{uv}}{k!}=\left(\exp\left(A\right)\right)_{uv}=\sum_{k=1}^{n}e^{\lambda_{k}}\mathbf{\mathbf{\psi}}_{k}(u)\mathbf{\psi}_{k}(v).
\end{equation}

It counts the total number of walks starting at node $u$ and ending
at node $v$, weighted in decreasing order of their length by a factor
$\frac{1}{k!}$. That is, the communicability function considers shorter
walks more influential than longer ones and penalize them appropriately
such that the whole series converges. The $G_{uu}$ terms of the communicability
function characterize the degree of participation of a node in all
subgraphs of the network, giving more weight to the smaller ones.
Thus, it is known as the subgraph centrality of the corresponding
node \cite{Subgraph_Centrality}. 

\section{Hyperspherical Embedding of Networks}

An important property of the communicability function of networks
is that it induces an embedding of the network into a given Euclidean
space. The important parameter in this case is the difference between
the number of weighted closed walks that start at (and return to) the corresponding
nodes $u$ and $v$, and the number of weighted walks that start at
node $u$ (respectively $v)$ and ends at the node $v$ (respectively
$u).$ This difference, which is defined below as $\xi_{uv}^{2}$ serves
as a quantification of the potential quality of communication channels
between two nodes. That is, if there are many routes that connect
nodes $u$ and $v$ together, and there are not many routes that starting
at the node $u$ (respectively $v)$ return to it, we can say that
most of ``information'' departing the node $u$ (respectively $v)$
with destination to the node $v$ (respectively $u)$ will arrive
at it. Thus, there is a potential good quality of communication between
these two nodes. The other way around is very clear as if there are
many returning routes to the nodes and very few connecting them, most
of the information departing one node will never arrive at the other.
Let us now define these terms formally. Based in the previous intuition
we define the following quantity:
\begin{equation}
\xi_{uv}^{2}=G_{uu}+G_{vv}-2G_{uv}
\end{equation}

Because $G=\exp\left(A\right)$ is positive define we can express
it as a Gram matrix of the form

\begin{equation}
G=X^{T}X,
\end{equation}
where $X=\left[\vec{x}_{1},\ldots,\vec{x}_{n}\right]$ and 

\begin{equation}
\vec{x}_{u}=\exp\left(\varLambda/2\right)\vec{\varphi}_{u},
\end{equation}

It is straighforward to realize that

\begin{equation}
\begin{array}{lll}
\vec{x}_{u}\cdot\vec{x}_{u} & =\left(\exp\left(\varLambda/2\right)\vec{\varphi}_{u}\right)^{T}\exp\left(\varLambda/2\right)\vec{\varphi}_{v}\\
 & =\vec{\varphi}_{u}^{T}\exp\left(\varLambda/2\right)\exp\left(\varLambda/2\right)\vec{\varphi}_{v}\\
 & =\vec{\varphi}_{u}^{T}\exp\left(\varLambda\right)\vec{\varphi}_{v}\\
 & =\sum_{k=1}^{n}e^{\lambda_{k}}\mathbf{\mathbf{\psi}}_{k}(u)\mathbf{\psi}_{k}(v)\\
 & =G_{uv.}
\end{array}
\end{equation}
Then, we can express $\xi_{uv}^{2}$ in terms of the vectors $\vec{x}_{u}$
and $\vec{x}_{u}$ as
\begin{equation}
\begin{array}{ll}
\xi_{uv}^{2} & =\vec{x}_{u}\cdot\vec{x}_{u}+\vec{x}_{v}\cdot\vec{x}_{v}-2\vec{x}_{u}\cdot\vec{x}_{v}\\
 & =\left\Vert \vec{x}_{u}-\vec{x}_{v}\right\Vert ^{2},
\end{array}
\end{equation}
which means that $\xi_{uv}^{2}$ is a Euclidean distance (metric)
between the corresponding nodes and that $\vec{x}_{u}$ is the position
vector of the corresponding node in such Euclidean space. Previously
we we have proved that such embedding space is an $n$-dimensional
sphere \cite{Hyperspherical_embedding}. That is, the communicability
distance $\xi_{uv}^{2}$ induces an embedding of the graph $\Gamma$ of
size $n$ into an $\left(n-1\right)$-sphere, of radius $R^{2}=\frac{1}{4}\left(c-\frac{\left(2-b\right)^{2}}{a}\right)$,
where $a=\vec{1}^{T}\exp\left(-A\right)\vec{1}$, $b=\vec{s}^{T}\exp\left(-A\right)\vec{1}$,
$c=\vec{s}^{T}\exp\left(-A\right)\vec{s}$, and $\vec{s}={\rm diag}(\exp\left(A\right))$.
The angles between the position vectors of the nodes $u$ and $v$
in the $\left(n-1\right)$-sphere are then given by \cite{Angles}
\begin{equation}
\theta_{uv}=\cos^{-1}\dfrac{G_{uv}}{\sqrt{G_{uu}G_{vv}}}.\label{eq:comm_angle}
\end{equation}

\begin{figure}[]
\begin{centering}
\includegraphics[width=0.6\textwidth]{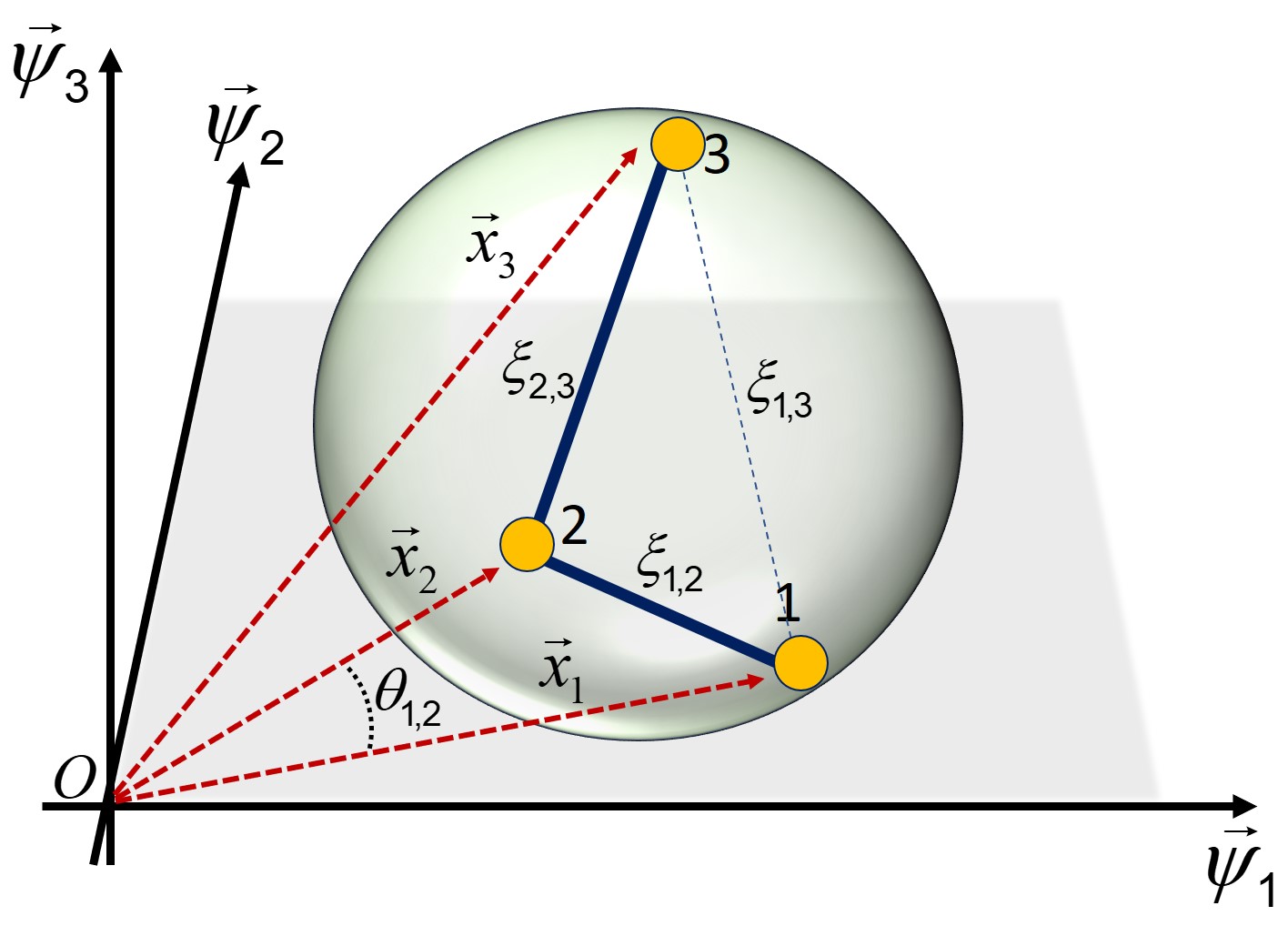}
\par\end{centering}
\caption{Illustration of the embedding of a simple graph with 3 nodes into
a 2-sphere. See main text for the description of the terms involved.}

\label{embedding}
\end{figure}

A pictorial representation of the communicability-induced geometry
of a graph is given in Figure \ref{embedding} for the case of a path
graph with three nodes $P_{3}$. In this case the embedding is realized
in a 3-dimensional sphere (2-sphere). The communicability distance
is the chord between the two corresponding nodes embedded in the sphere
and the communicability angle is the one formed between the centre
of coordinates and the two corresponding nodes.

The communicability angle is defined as the ratio between the weighted
number of walks connecting the two nodes to the number of walks starting
and ending at the same node. Thus, it also quantifies the goodness
of communication channels between the two nodes. However, while the
communicability distance is unbounded \textendash it can take an infinite
value for infinitely large graphs \textendash the communicability angle
is bounded as 
$
0^{\circ}\leq\theta_{uv}\leq90^{\circ}
$
\cite{Angles},
where the lower bound is reached for pairs of nodes communicating
with exceptionally good quality and the upper bound is reached for
pairs of nodes communicating in a very bad way. The communicability
angle is then a way of quantifying how efficiently a network uses
the communication space available to it. \textcolor{black}{For instance,
the average communicability angles for the connected graphs with four
nodes are: Path ($59.09^{\circ})$, Star ($55.37^{\circ})$, Cycle ($45.11^{\circ})$, triangle
with a pendant node ($44.97^{\circ})$, and complete graph ($21.48^{\circ})$. Clearly, the path graph is
the least efficient one, followed by the star\textemdash which are
the only two trees in this set\textemdash , then the graph consisting
of a triangle and a pendant node, the cycle and finally the complete
graph. This order agrees very well with our intuition corresponding
to the average communication efficiency of these graphs. For all the
previous reasons hereafter we concentrate our study on the communicability
angles instead of the communicability distances.}

\section{Nonmetric Multidimensional Scaling and Visualization}

The hyperspherical embedding induced by the communicability geometry of a network does not allow to visualize the corresponding network due to the high dimensionality of the embedding spaces. Then, we aim here to reduce such space dimensionality to a 3-dimensional (3D) Euclidean
space which allow us to visualize the network structure. We selected
the 3D space as it represents the largest dimensionality that we can
visualize with the minimum loss of information. There are several
techniques that can be used for this dimensionality reduction \cite{MDS}.
Here we selected the nonmetric multidimensional scaling (NMDS) mainly
due to the following reasons. In NMDS \cite{NMDS_1,NMDS_2} we select
a priori the number of axis that we wish\textendash 3 in this case\textendash instead
of the many axis selected by most of ordination methods. Also NMDS
is a numerical method that iteratively searches for a solution stopping
when an acceptable solution (or a predetermined number of iterations). NMDS is not an eigen-method like PCA or correspondence
analysis in which axis explain the variance of the data in a given
order according to the magnitude of the corresponding eigenvalue.
Finally, NMDS makes little assumptions about the data points which
makes it very suitable for a wide variety of data. A clear disadvantage
of NMDS with respect to other ordination methods is that its solution
is not unique due to its numerical nature. However, the current availability
of more powerful computational resources allows the execution of several
realizations with different initial conditions, to select the best
possible solution in a way that avoids getting trapped in local minima due
to the use of an heuristic algorithm. In general, multidimensional
scaling can be defined as a \emph{method that represents measurements
of similarity (or dissimilarity) among pairs of objects as distances
among points of a low-dimensional multidimensional space}'' \cite{MDS}.

Here we consider the matrix $\theta=\left[\theta_{pq}\right]_{n\times n}$
as our similarity matrix. That is two nodes of a graph/network are
similar if they have communicability angles close to $0^{\circ}$
and dissimilar if they have an angle close to $90^{\circ}$. We then transform this matrix as described in the Supplementary Information to directly apply NMDS. 

In order to avoid getting trapped into local minima, we make a series
of replications, with different initial random configurations, and
we select the ordination with the best fit. In addition of this nonmetric
scaling for the dissimilarities we have also implemented metric scaling
ones. In particular we consider the following four metric scaling:
'metricstress', which calculates the stress, normalized with the sum
of squares of the dissimilarities, 'metricsstress', which calculates
the squared stress, normalized with the sum of 4th powers of the dissimilarities;
'sammon', which obtains Sammon's nonlinear mapping criterion, and
'strain', which is a criterion equivalent to that used in classical
multidimensional scaling. All of these metrics are implemented in
Matlab\textregistered. 

As we have different metrics we will obtain different ``best'' reduced
angle matrices, one for each of scaling used. We noticed that the
root mean square error (RMSE) between the $n$-dimensional angle matrix
and the reduced 3D one was fooled by those angles which were relatively
close in both matrices although there were angles which differed significantly
from each other in the two matrices (see further example). Instead
of using RMSE as comparison method, we propose a spectral comparison
between the two matrices. That is, we calculated the eigen-distance
between $\theta$ and $\hat{\theta}$ (estimated angles in the 3D
space) from their eigenvalues. In this way we compute the sum of the
squares of the difference between the eigenvalues of both matrices
as our measure of quality of the fit to compare the different methods. In Table \ref{matrices} we give an example of the use of NMDS for
reducing the dimensionality of the communicability angle matrix to
a three-dimensional space. 

\begin{figure}[]
\begin{centering}
\subfloat[]{\includegraphics[width=0.4\textwidth]{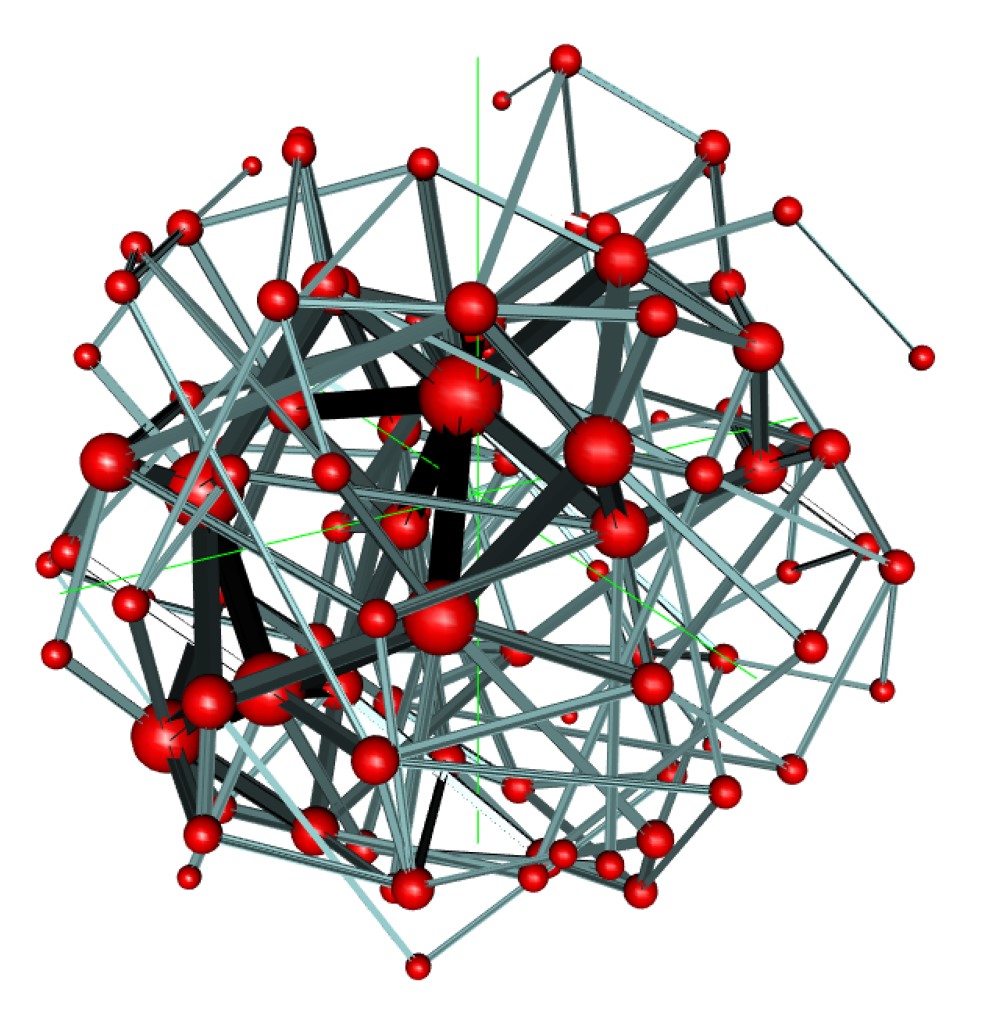}}\subfloat[]{\includegraphics[width=0.4\textwidth]{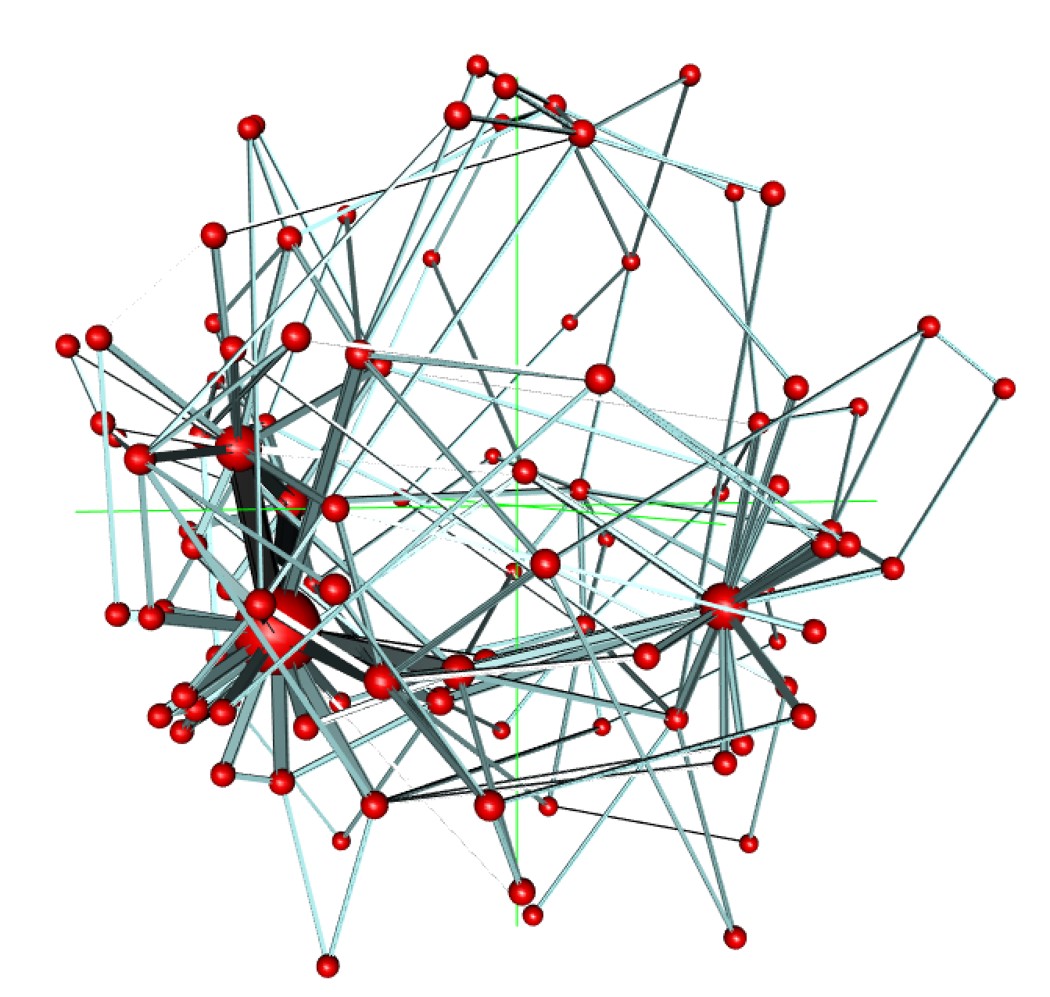}}
\par\end{centering}
\begin{centering}
\subfloat[]{\includegraphics[width=0.38\textwidth]{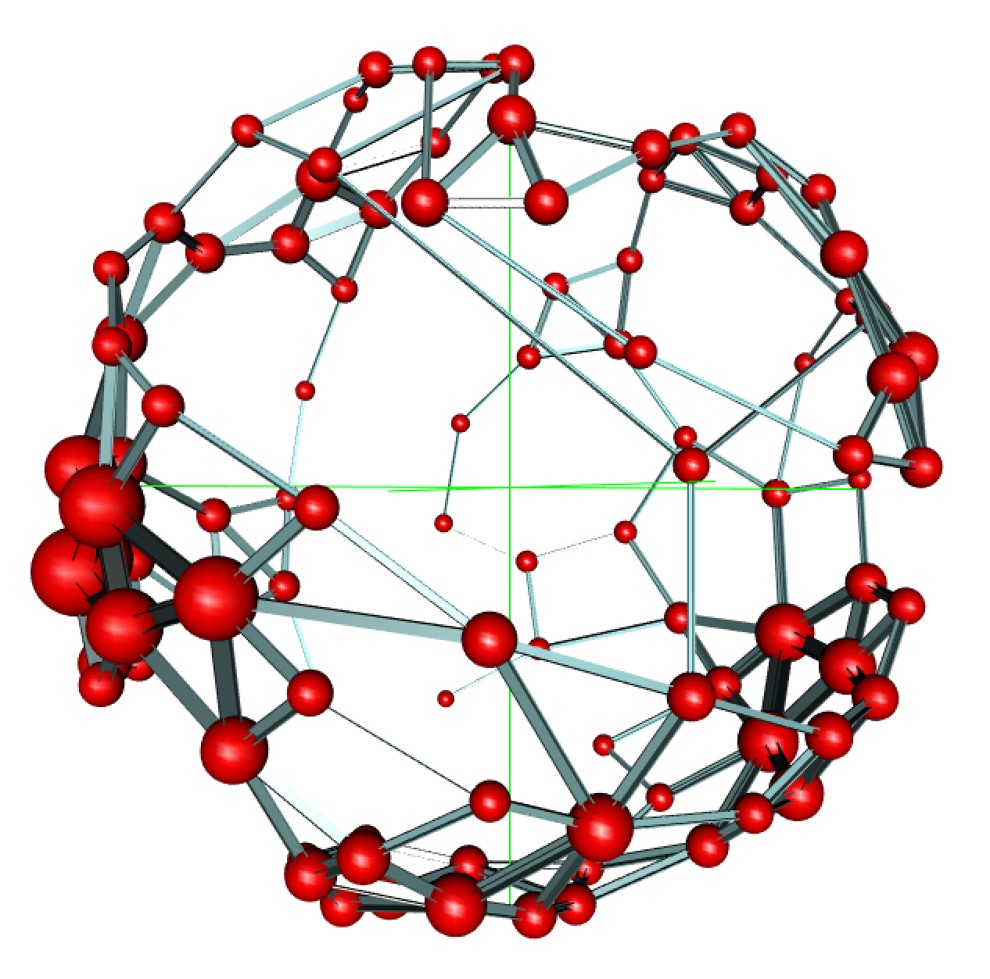}}\subfloat[]{\includegraphics[width=0.4\textwidth]{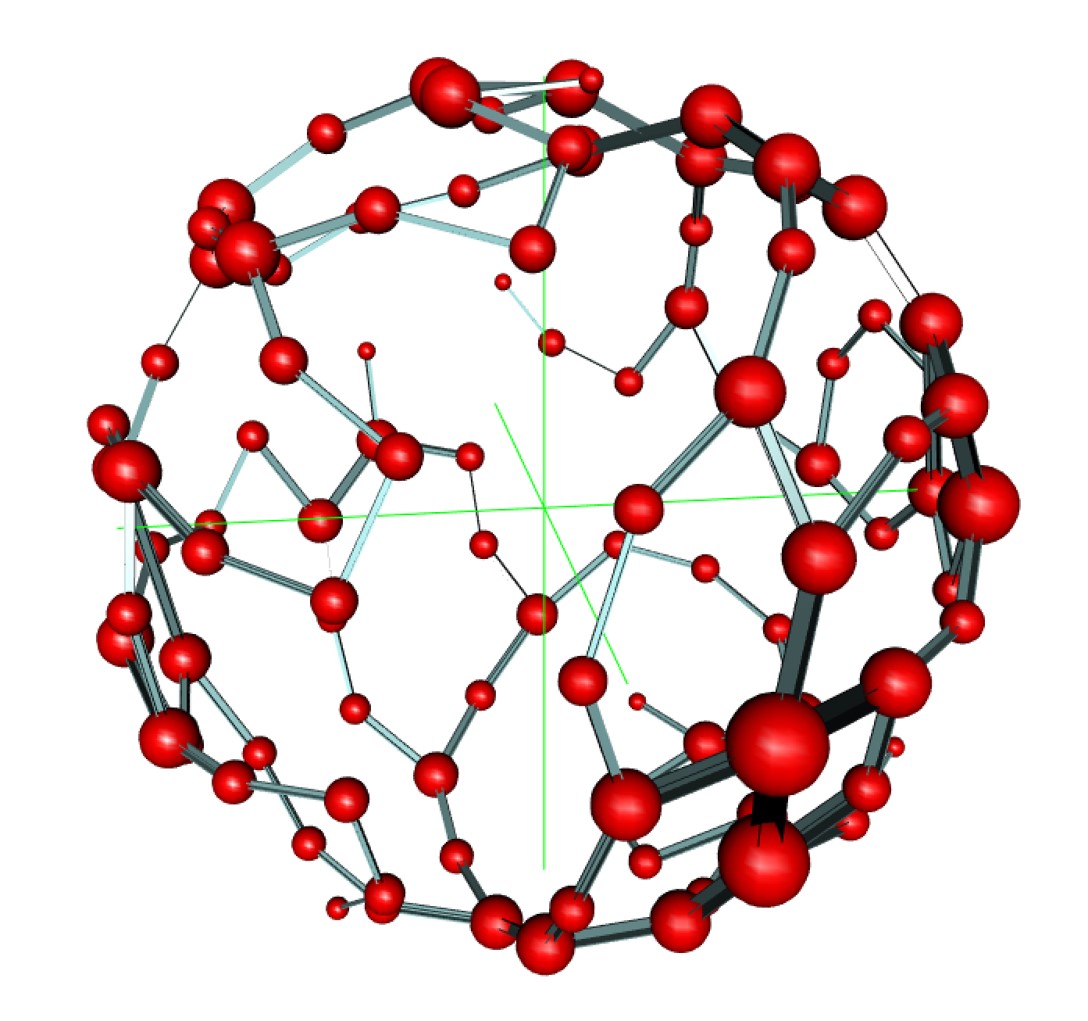}}
\par\end{centering}
\caption{Illustration of the plot of random graphs on the reduced 3D space
of communicability angles obtained by nonmetric multidimensional scaling.
(a) Erd\H{o}s-R\'enyi (ER) graph with $n=100$, and $m=200$, (b)
Barab\'asi-Albert (BA) graph with $n=100$, and $m=191$, (c) spatial
Gabriel graph with $n=100$, and $m=118$ and (d) Relative Neighborhood
Graph (RNG) with $n=100$, and $m=118$. The two spatial graphs in
(c) and (d) are special cases of $\beta$-skeleton graphs with $\beta=1$
and $\beta=2$, respectively. }
\label{random}
\end{figure}

We then consider some random networks based on three different construction
methods. The first is based on the Erd\H{o}s-R\'enyi (ER) model \cite{ER_model},
the second is based on the Barab\'asi-Albert (BA) model \cite{BA_model}
and the two others are based on the $\beta$-skeleton method for generating
spatial graphs \cite{Beta_skeleton,Beta_2}, in particular we study
the Gabriel graph ($\beta=1$) \cite{Gabriel} and the relative neighborhood
graph ($\beta=2$) \cite{Beta_skeleton}. In all cases the networks
constructed have 100 nodes and the number of edges depended on the
type of method used, e.g. 200 for ER, 191 for BA and 118 for the spatial
networks. In Figure\ref{random} we illustrate the 3D representation
of these four networks after the reduction of the dimensionality of
the communicability hyperspace using the multidimensional scaling method.
The best scaling were produced by Sammon (ER, Gabriel and RNG) and
by metricstress (BA) and the values of SE are: 347.97 (ER), 366.06
(BA), 348.02 (Gabriel) and 313.88 (RNG). 

\section{Cluster Analysis}
In our context of network analysis the problem of clustering in the
multimensional communicability space consists in having nodes close
to each other if they share certain structural similarities which
make them to cluster together, while those structural dissimilar nodes
are placed far apart in the 3D embedding studied here. The problem
of clustering is one of the most popular tasks in machine learning
and data science in general. Its main goal is to divide a set of objects
into clusters in an unsupervised manner, such that those objects in
a cluster are similar to each other and somehow dissimilar to those
on other clusters. In the context of network analysis this problem
is mainly studied for the detection of the so-called network communities.
We point the reader to \cite{Network_communities_1,Network_communities_1.5,Network_communities_2,Network_community_3}
for overviews on the most used community detection techniques in networks.
Most of these methods are defined on the basis of the internal density
of links in the clusters and its relation to the inter-cluster density
of links. Although this density-based definition is theoretically
appealing it leaves a lot of other possibilities outside its scope (see introduction).
In contrast, here we define a cluster as a homogeneous group of nodes
in terms of their communicability angles in the multidimensional communicability
hypersphere in which the network is embedded. We know that a characteristic
feature of multidimensional scaling is that it produces an embedding
of the data points into a reduced space such that the points which
are located close to each other are similar according to some empirical
characteristic, and those that are far apart are different. Thus,
we also analyze the clustering problem of nodes in networks using
the reduced 3D space produced by the multidimensional scaling studied
in this work. In both cases, we expect that the nodes inside a cluster
are homogeneous in terms of their total or ``reduced'' communicability
angles, and nodes in different clusters are expected to be heterogeneous
in this respect. These clusters may be or may be not related to density-based
communities, but in any case we have a clear an unambiguous definition
of them.

In this work we aim at finding communicability clusters in the sense of K-Means \cite{Clustering_3,K-means}, i.e., we propose a partitional clustering method \cite{Network_communities_1.5} using the embeding
induced by the communicability function. Our selection of this method
is based on its great popularity for detecting clusters in an unsupervised
manner, which have produced a large number of documented examples
of applications in areas such as cheminformatics, bioinformatics,
data mining, natural language processing, among others. We also know
the main drawbacks of this method, mainly the fact that K-Means does
not guarantee a convergence to a global minimum, with its final clusters
heavily dependent on the initial centroids \ref{SI_kmeans}. 

Due to the fact that we have to pre-select the number of clusters
$K$ we need to compare the quality of the partition for the different
values of the number of clusters selected. With this goal we use several
cluster validity indexes (CVIs) for estimating $K$ \cite{Clustering_validation_1,Clustering_validation_2}. Here we use three popular CVIs, namely the Calinski-Harabasz index \cite{Validation_CH}, the Silhouette index \cite{Validation_S}, and the Davies-Bouldin index \cite{Validation_DB}, which are described in the Supplementary Information accompanying this paper.
%
%
\subsection{Resolution}
\begin{figure}[]
\begin{centering}
\subfloat[]{\includegraphics[width=0.32\textwidth]{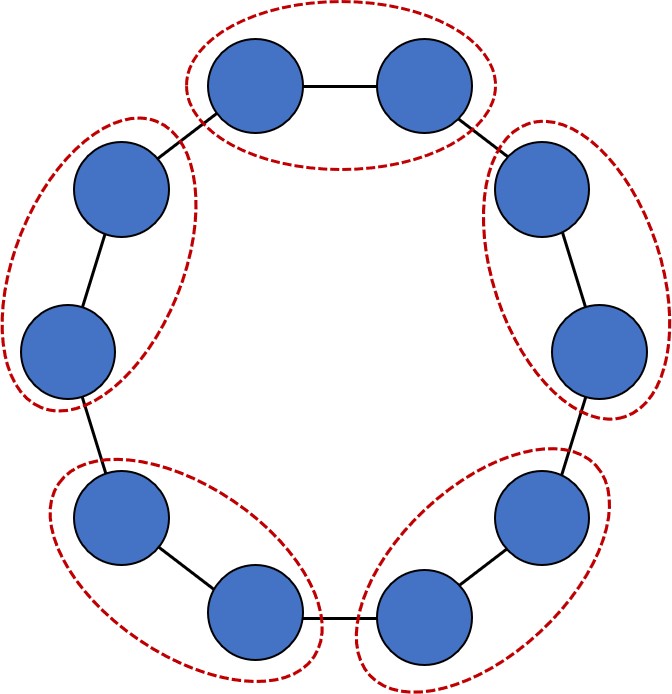}

}\subfloat[]{\includegraphics[width=0.32\textwidth]{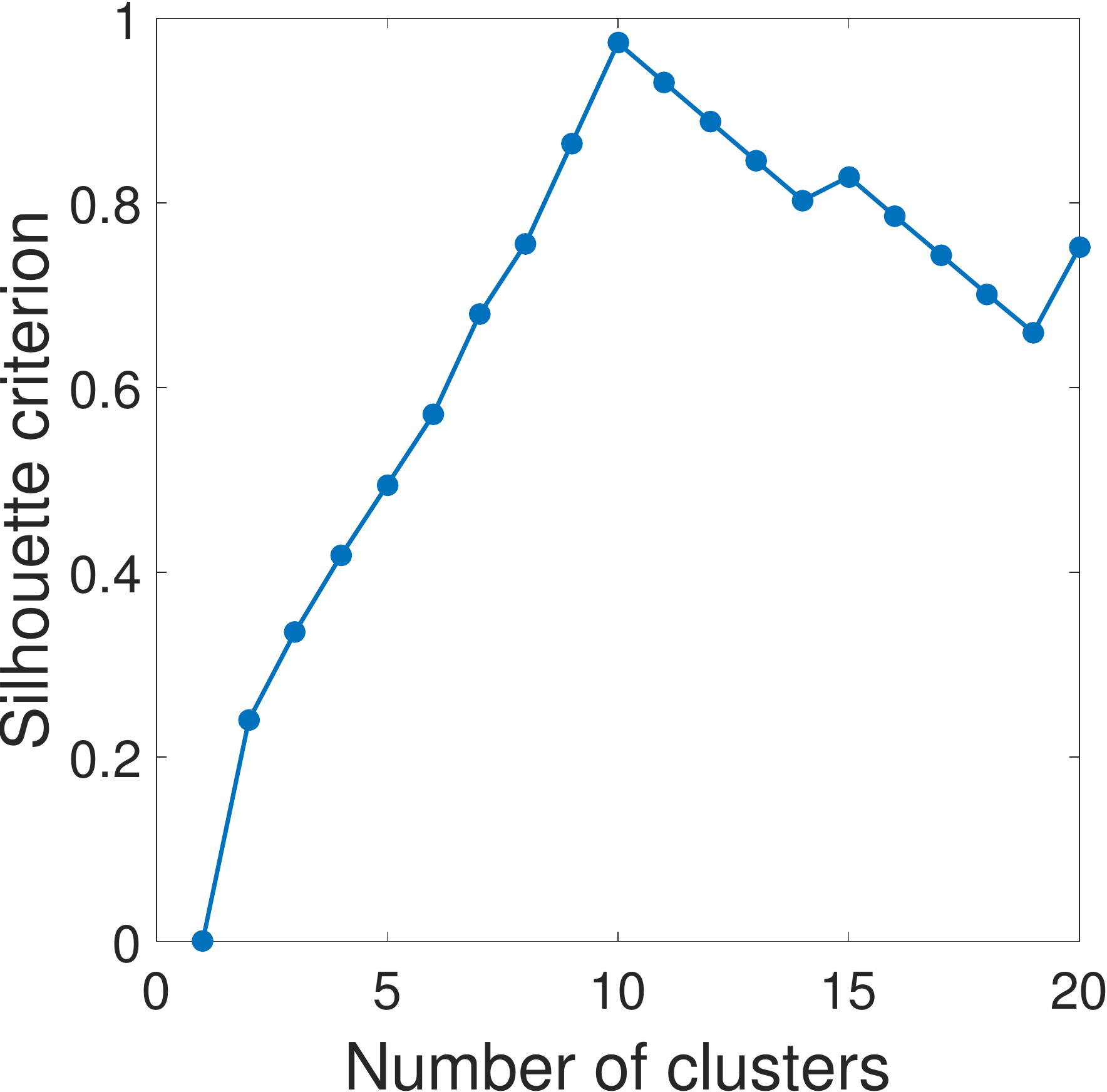}

}\subfloat[]{\includegraphics[width=0.32\textwidth]{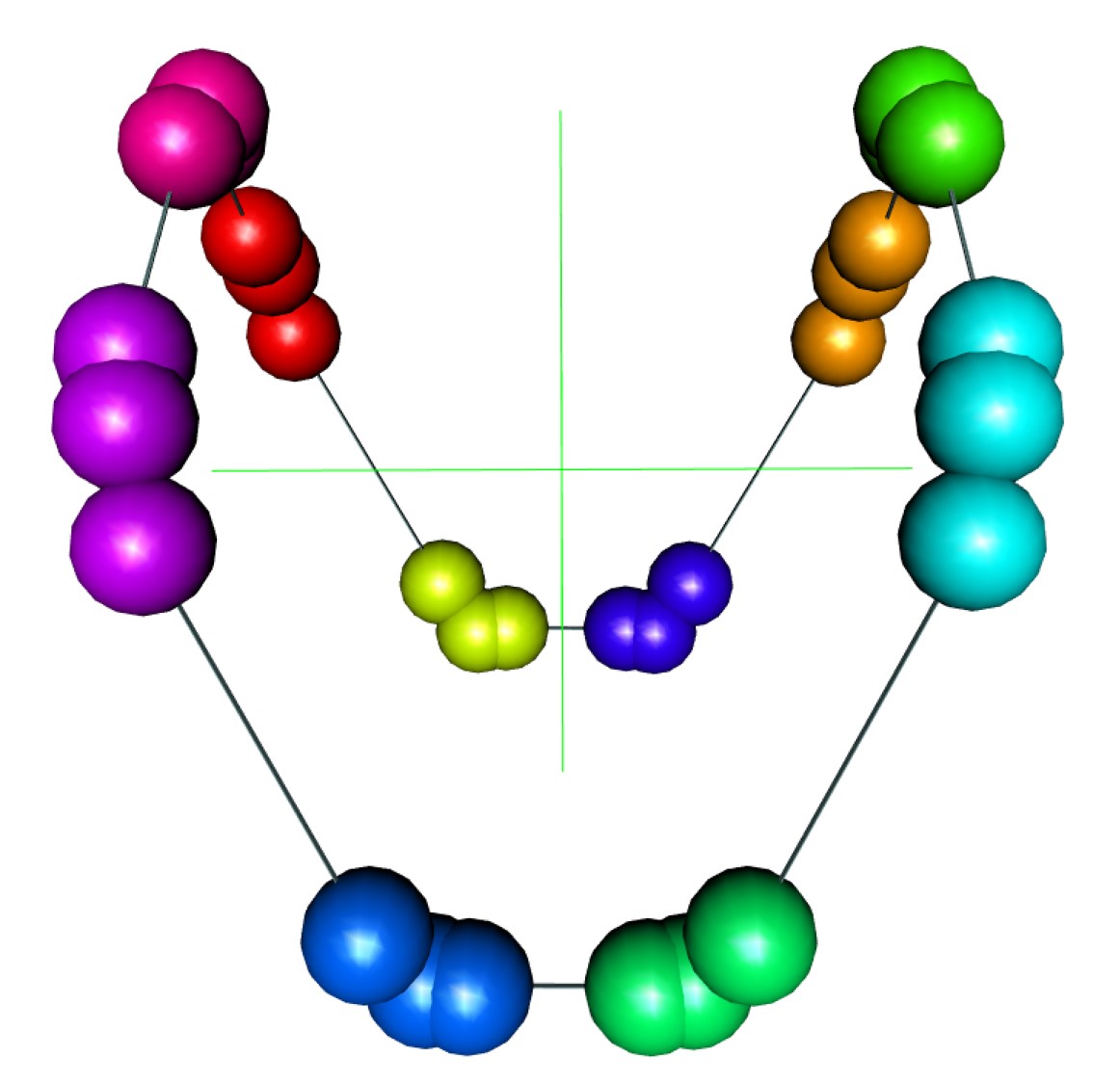}

}
\par\end{centering}
\caption{(a) Illustration of a caveman graph in which modularity-based methods
fail to detect the correct number of communities due to their merging
of pairs of close clusters (clusters of nodes are represented as blue
circles). (b) Plot of the variation of the Silhouette index for the
different number of clusters in the graph illustrated in (a). Notice
that the best performance is obtained for 10 clusters which is the
correct number. (c) Illustration of the clusters obtained from the
full-dimensional communicability angles and K-Means using Silhouette.
The plot is obtained in the reduced 3D communicability space obtained from nonmetric multidimensional scaling.}
\label{REsolution}
\end{figure}

\begin{figure}[]
\begin{centering}
\includegraphics[width=0.75\textwidth]{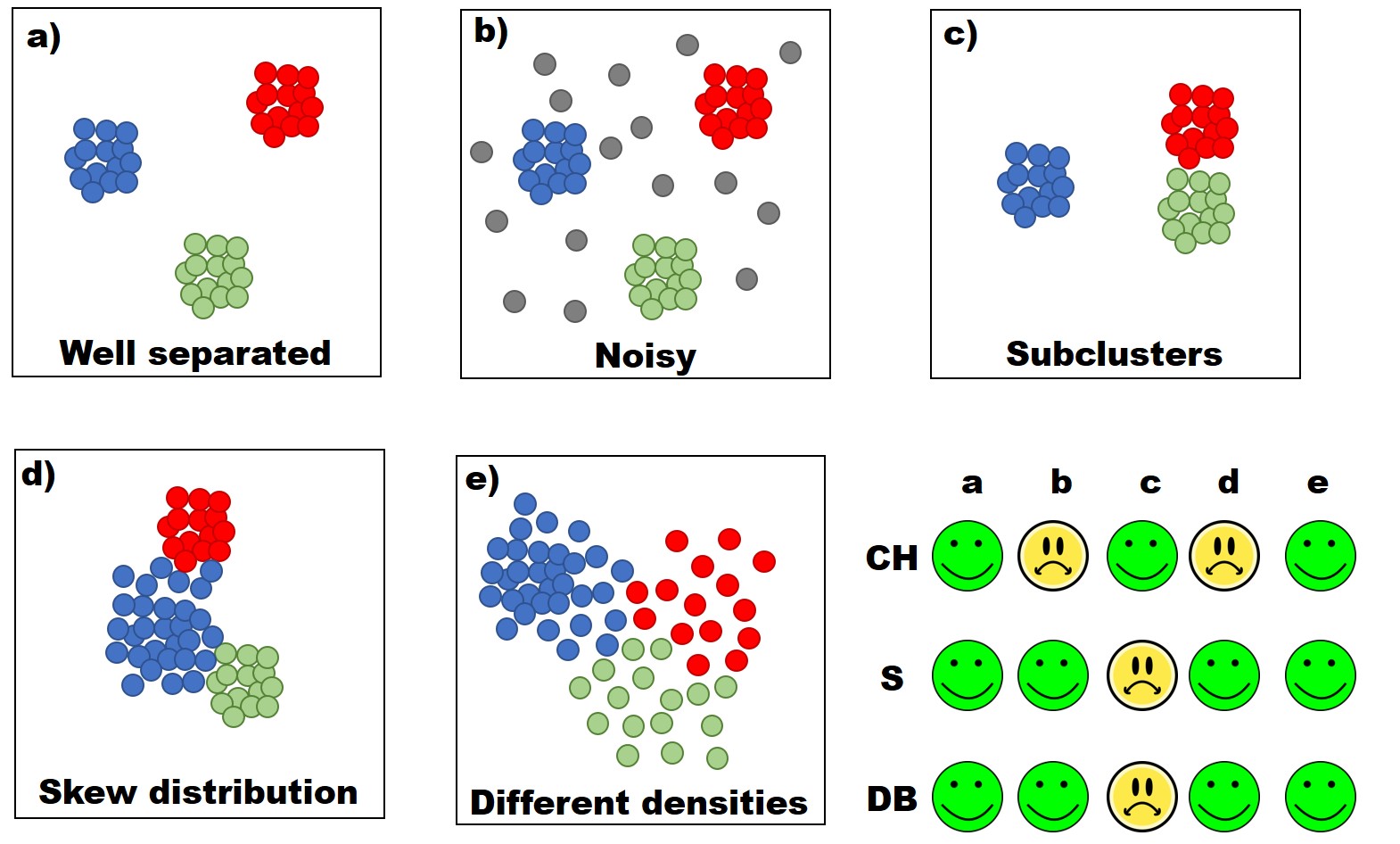}
\par\end{centering}
\caption{Illustration of the impact of several cluster properties on the three
CVIs used in this work according to Liu et al. The CVIs under consideration
are \textit{Calinski-Harabasz (CH) index, Silhouette (S) index }\textcolor{black}{and
}\textit{Davies-Bouldin (DB) index.}\textcolor{black}{{} In the last
panel a smiling face indicates that the method performs well on this
situation and a sad face indicates that the method fails\textendash under-
or overestimate the number of clusters\textendash on this situation.}}
\label{clusters_problems}
\end{figure}

One of the most important problems that have been detected when studying
clustering problems on networks is that of the so-called resolution
limit \cite{Resolution_limit}. The problem is simply described as
the gluing of small clusters produced by some clustering methods\textendash particularly
those based on modularity\textendash when there are many of such small
highly connected clusters which are loosely connected among them.
The paradigmatic example of the resolution limit is given by the caveman
network illustrated in Fig.\ref{REsolution} (a) in which every solid
circle represents a clique. In this type of situation, modularity-based
methods detect half the number of clusters existing in the network
due to the merging of clusters by pairs as indicated in Fig.\ref{REsolution}
(a) by the broken red ellipses. We have tested the current method
for the existence of such resolution limit problem and the results
are illustrated in Fig.\ref{REsolution} (b) and (c) for the use
of the Silhouette method using the full-dimensional communicability
angle matrix. The results are exactly the same for the reduced communicability
angle matrix as well as for the three different CVIs used here. As
can be seen the Silhouette index has a clear maximum peak at $K=10$
indicating the existence of the 10 expected communities, each of them
formed by a clique of 5 nodes. In Fig.\ref{REsolution} (c) we illustrate
the plot of the communities in the reduced 3D space using the nonmetric
multidimensional scaling method previously described. It is easy to
see that each cluster is well separated from each other in the graphic
and that the nodes in each clique are grouped together in a close
space.
The resolution problem is only one of the several situations that
we can find when studying clustering of points. A group of these situations
have been studied by Liu et al. \cite{Clustering_validation_2} and
they include, apart from the desired case of well-separated clusters,
the existence of noisy clusters, subclusters, clusters distributed
in a skew way and the presence of clusters of different densities
(see Fig.\ref{clusters_problems} for illustration). In Fig.\ref{clusters_problems}
(f) we illustrate the results of Liu et al.'s \cite{Clustering_validation_2}
experiments with the K-means method and the three CVIs used here.
As can be seen CH fails in identifying correctly the existing clusters
when they are noisy, and S and DB fail when there are subclusters
as indicated in Fig.\ref{clusters_problems} (c). According to the
results of Liu et al. \cite{Clustering_validation_2} in the presence
of noise or clusters with skewed distributions CH tends to predict
more clusters than the truly existing ones. On the other hand, when
there are subclusters, both S and BD tend to predict less clusters
than the existing ones, due to the fact that they merge together small
clusters. Then, the methods used here are somehow complimentary to
each other and we should have in mind the situations where they clearly
fail when analyzing the real-world datasets that will be studied in
the next section. 
\subsection{Real-world networks with ground truth}
Now we turn our attention to some real-world networks for which a
sort of ``ground-truth'' for their partition into clusters exists.
These networks constitute a benchmark for testing algorithms used
for community detection in networks. Thus, we use them here to test
the three CVIs used in the current work by using first the full-dimensional
communicability angles matrices and the K-means method of clustering.
In selecting the ``best'' approach we use the \textit{\textcolor{black}{Normalized
Mutual Information}} (NMI), as defined by Strehl and Ghosh \cite{NMI} (see Supplementary Information), to measure the performance of the different
algorithms on detecting clustering in comparison with the ground-truth.
In addition to NMI we also report the \textit{\textcolor{black}{modularity
index}} for the partitions found as it is usually reported for network communities analysis \cite{modularity}. 
We then apply our clustering algorithm to the following networks:
the karate club of Zachary \cite{karate_club}, the dolphin social
network of Lusseau et al. \cite{dolphins}, the network of political
weblogs studied by Adamic and Glance \cite{plo_blogs} the network
of books about politics studied by Krebs (see \cite{modularity}),
and the network of regular season games between Division I-A college
football teams in the year 2000 \cite{Girvan_Newman}. The results
obtained with the full-dimensional communicability angle matrices
and K-means are given in Table \ref{Results_real}. The results produced
with the reduced 3D communicability angles space do not improve in
any case those obtained with the full space and are not reported here.
The main reason for that decision is that since information is lost
in the dimensionality reduction process. it would be nonsense to apply
the clustering algorithm to those if computational power allows to
analyze the full communicability angles space data.
\begin{table}[]
\centering
\resizebox{\textwidth}{!}{
\begin{tabular}{llllllllllllll}
\hline
         & \multicolumn{2}{l}{ground truth} & \multicolumn{3}{l}{Silhouette}             & \multicolumn{3}{l}{Calinski-Harabasz}      & \multicolumn{3}{l}{Davies-Bouldin}         & \multicolumn{2}{l}{Methods} \\ \cline{2-14} 
network  & C              & Q               & C          & NMI           & Q             & C          & NMI           & Q             & C          & NMI           & Q             & NMI          & Q            \\ \hline
Karate   & 2              & 0.37            & \textbf{2} & \textbf{1.00} & \textbf{0.37} & \textbf{2} & \textbf{1.00} & \textbf{0.37} & \textbf{2} & \textbf{1.00} & \textbf{0.37} & 1.00         & 0.37         \\
Dolphins & 2              & 0.38            & \textbf{2} & \textbf{0.89} & \textbf{0.38} & \textbf{2} & \textbf{0.89} & \textbf{0.38} & \textbf{2} & \textbf{0.89} & \textbf{0.38} & 0.89         & 0.38         \\
Football & 12             & 0.55            & 11         & 0.91          & 0.66          & 12         & 0.91          & 0.64          & 13         & 0.90          & 0.66          & 0.91         & 0.66         \\
PolBooks & 2              & 0.41            & \textbf{2} & \textbf{0.61} & \textbf{0.44} & 7          & 0.58          & 0.45          & \textbf{2} & \textbf{0.61} & \textbf{0.44} & 0.60         & 0.44         \\
PolBlogs & 2              & 0.41            & \textbf{2} & \textbf{0.72} & \textbf{0.52} & -          & -             & -             & \textbf{2} & \textbf{0.72} & \textbf{0.52} & 0.71         & 0.52         \\ \hline
Average  &                &                 &            & 0.83          &               &            &               &               &            & 0.82          &               &              &              \\ \hline
\end{tabular}
}
\caption{Modularity $Q$ and $NMI$ for the 5 stardard networks used for community
detection using communicability angle and four different statistical
clustering methods. C: number of clusters in the ground truth and
detected for each community detection method. In bold cases where
the number of clusters found by the methods coincide with the number
of clusters present in the ground truth.}
\label{Results_real}
\end{table}
We now compare these results with those reported in the literature
using five different ``popular'' methods for community detection.
These methods are: Louvain, FastGreedy, Infomap, Eigenvector and LP
(for description of these methods and references see \cite{Network_communities_1.5}).
In Table \ref{Results_real_reported} we give the values of the $NMI$
and $Q$ for the networks studied here. In addition we report the
average value of $NMI$ obtained for each network as well as the average
of this parameter for every method. As can be seen Louvain, FastGreedy,
Infomap and Eigenvector produce $NMI$ in the range of 0.62-0.64 for
all the five networks studied, and LP produces 0.71. \textcolor{black}{Our
method using of the three CVIs studied here produces superior results
for these 5 networks, with the exception of CH that fails to find
clusters for the network of political blogs. For this network CH always
predicts the highest possible number of clusters according to the
range of values of $K$ introduced in the K-means method. As we have
seen before }in the presence of noise or clusters with skewed distributions
CH, tends to predict more clusters than the truly existing ones. In
terms of the individual networks, our method produces the best results
for the networks of karate club and dolphins using any of the three
CVIs. For the network of football our method based on Silhouette produces
similar results to the best performances obtained by Infomap and LP.
For Political books and political blogs our results using Silhouette
and Davies-Bouldin are far better than any of the five standard methods
used for comparison. Thus, in general our method using either S or
DB methods produces high standard clusters according to the metrics
used here and for the set of standard datasets considered.
\begin{table}[]
\centering
\resizebox{\textwidth}{!}{
\begin{tabular}{llllllllllll}
\cline{1-12}
         & \multicolumn{2}{c}{Louvain}     & \multicolumn{2}{c}{FastGreedy} & \multicolumn{2}{c}{Infomap} & \multicolumn{2}{c}{Eigenvector} & \multicolumn{2}{c}{LP} & \multicolumn{1}{c}{\multirow{2}{*}{NMI}} \\ \cline{1-11}
network  & NMI                      & Q    & NMI            & Q             & NMI          & Q            & NMI            & Q              & NMI        & Q         & \multicolumn{1}{c}{}                     \\ \hline
Karate   & 0.59                     & 0.42 & 0.69           & 0.38          & 0.70         & 0.40         & 0.68           & 0.39           & 0.70       & 0.40      & 0.73                                     \\
Dolphins & 0.48                     & 0.52 & 0.61           & 0.50          & 0.50         & 0.52         & 0.54           & 0.49           & 0.69       & 0.50      & 0.64                                     \\
Football & 0.88                     & 0.60 & 0.70           & 0.55          & 0.92         & 0.60         & 0.70           & 0.49           & 0.92       & 0.60      & 0.83                                     \\
PolBooks & 0.51                     & 0.52 & 0.53           & 0.50          & 0.49         & 0.52         & 0.52           & 0.47           & 0.57       & 0.50      & 0.54                                     \\
PolBlogs & 0.63                     & 0.43 & 0.65           & 0.43          & 0.48         & 0.42         & 0.69           & 0.42           & 0.69       & 0.43      & 0.64                                     \\ \hline
Average  & \multicolumn{1}{c}{0.62} &      & 0.64           &               & 0.62         &              & 0.63           &                & 0.71       &           &                                          \\ \hline
\end{tabular}
}
\caption{Modularity $Q$ and $NMI$ for the 5 standard networks used for community
detection using 6 different methods from the literature. The average
value of the $NMI$ for each method is given as the last row of the
Table. The average $NMI$ obtained for each of the networks using
the 6 methods considered is given as the last column of the Table.}
\label{Results_real_reported}
\end{table}
Finally, we illustrate the clusters obtained by the best CVI used
before in K-Means for four of the five networks studied. In order
to make these graphics we have projected the community structure onto
the reduced 3D communicability space obtained by the nonmetric multidimensional
scaling previously described. As can be seen there is a good separation
of the clusters obtained that can be easily visualized in these reduced
spaces. 

\begin{figure}[]
\centering
\subfloat[]{\includegraphics[width=0.45\textwidth]{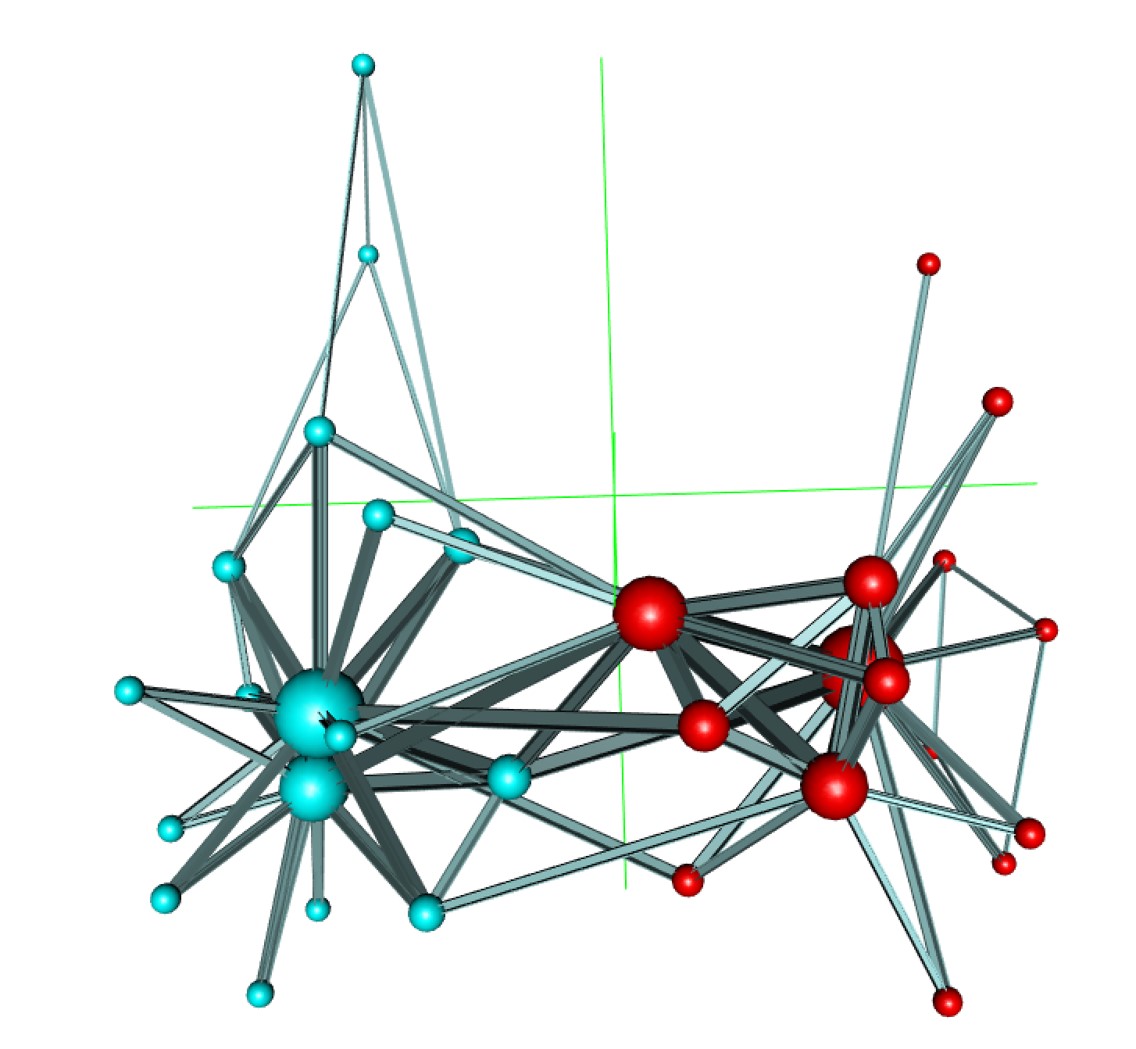}}\subfloat[]{\includegraphics[width=0.45\textwidth]{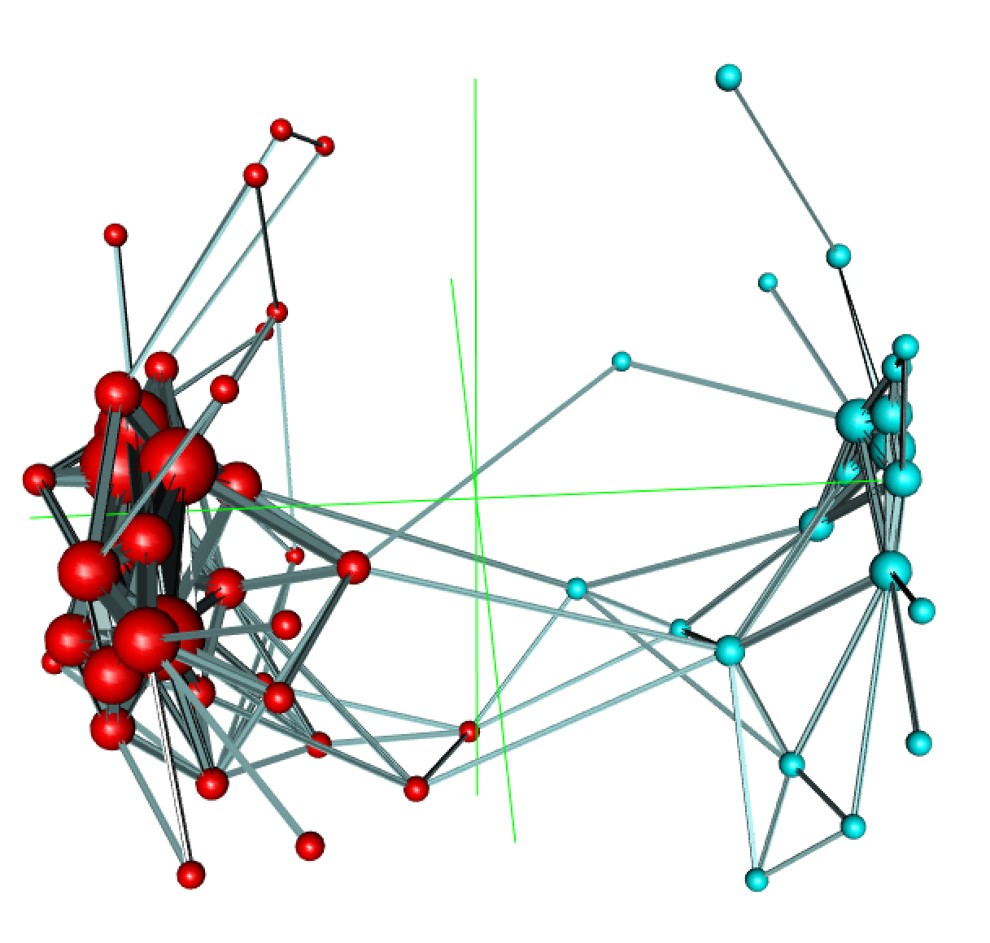}}

\subfloat[]{\includegraphics[width=0.45\textwidth]{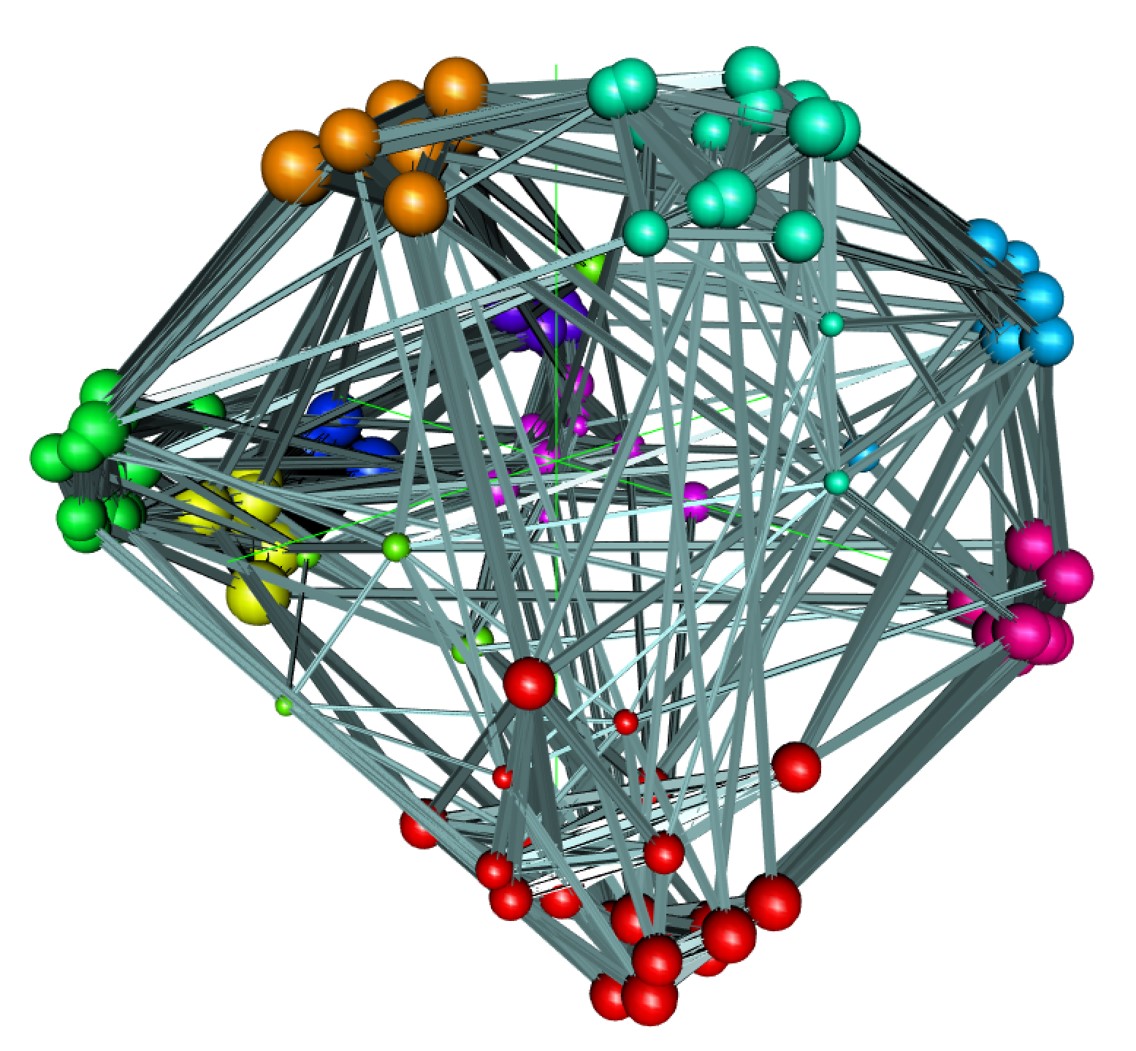}}\subfloat[]{\includegraphics[width=0.45\textwidth]{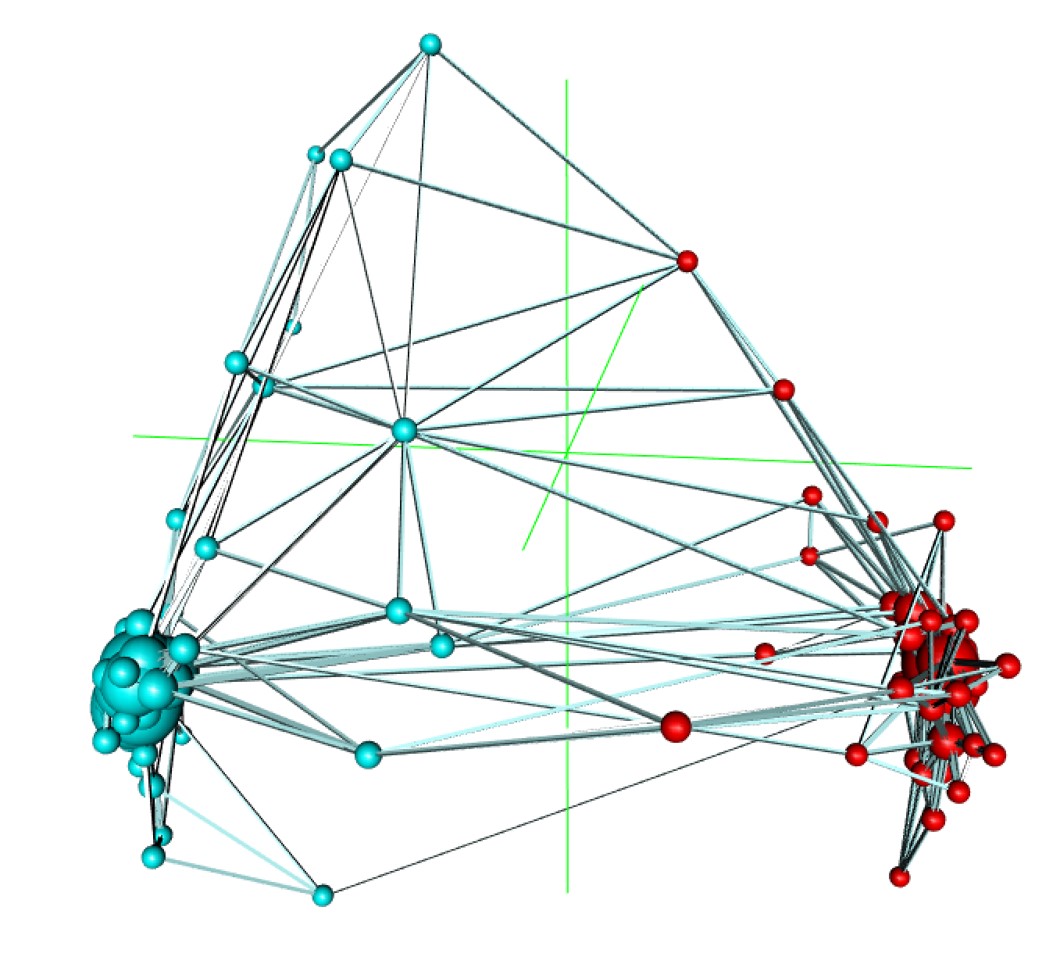}}
\caption{Illustration of the clusters obtained by K-Means for the networks
of karate club, dolphins, football and political books. The CVIs used
for the detection of the best partition are described in the main
text. The networks are embedded into the reduced 3D space obtained
by nonmetric multidimensional scaling. }
\end{figure}
\subsection{Unsupervised clustering without ground truth}
\subsubsection{Analysis of a citation network}
Here we consider a citation network in which nodes represent papers that cite Milgram's
1967 \textit{\textcolor{black}{Psychology Today}} paper or use ``\textit{\textcolor{black}{Small
World}}'' in title, collected until July 23, 2002 \cite{Pajek_dataset}.
Two nodes are connected if one of the papers cites the other. Although
the network is directed we have symmetrized it for the current analysis.
In this scenario we would like to investigate all the possible clusters
existing in that network, which may represent group of papers coming
from closely related scientific communities. Thus, our goal is to
find the finest grain structure of such scientific communities in
the network. 

\begin{figure}[]
\begin{centering}
\includegraphics[width=0.6\textwidth]{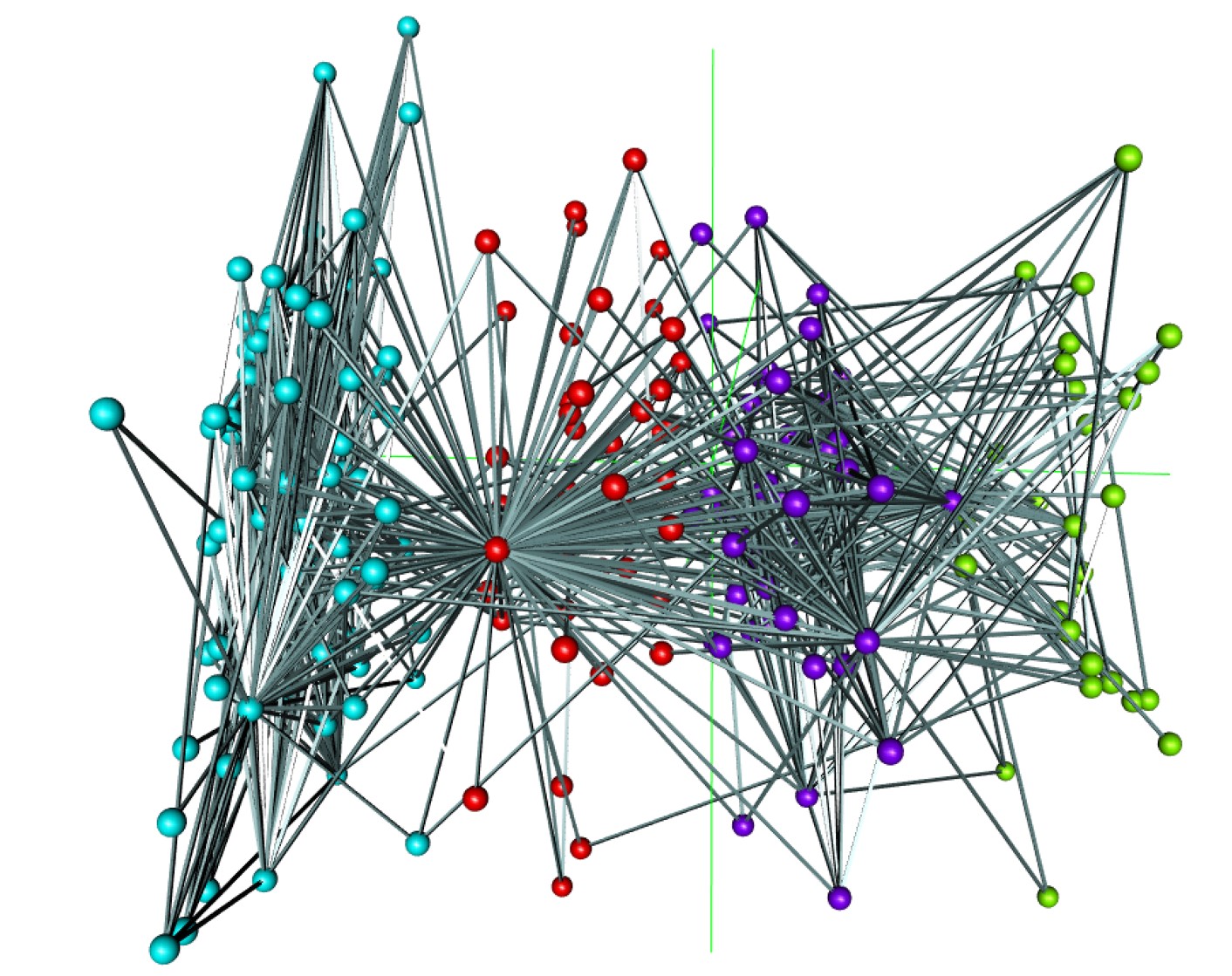}
\par\end{centering}
\caption{Illustration of the 3D embedding of the citation network of ``Small-World''
in which papers that cite Milgram's 1967 Psychology Today paper or
use Small World in title, until July 23, 2002 are accounted. Two nodes
are connected if one of the papers cites the other. Although the network
is directed we have symmetrized it for the current analysis. }
\label{Small_World}
\end{figure}

In this case Silhouette identifies 4 communicability clusters, Davies-Bouldin
identifies 3 clusters and Calinski-Harabasz fails to identify any
clustering as the index always increase with the maximum number of
clusters to be identified. Both S and DB identifies a large cluster
composed by 108 papers which mainly consists of papers published in
Physics or multidisciplinary journals. DB includes 7 papers not included
by S in this cluster, which are mainly from exogenous areas to the
main subject in this cluster. They includes apart from a paper in
the \textit{\textcolor{black}{European Physics Journal}}, papers published
in \textit{\textcolor{black}{Ethnology}}, \textit{\textcolor{black}{American
Sociological Review}}, \textit{\textcolor{black}{IEEE Internet Computing}},
\textit{\textcolor{black}{Society}} and an \textit{\textcolor{black}{ACM}}
conference. The cluster identified by Silhouette is colored in cyan
in Fig.\ref{Small_World}. The second cluster identified by DB is
identical to the one identified by S and consists of 45 papers, which
include Milgram's itself (see red nodes in Fig. \ref{Small_World}).
This cluster mainly consists of papers in the area of quantitative
social sciences and include papers about modifications of the ``small-world''
model, quantification of clustering, social distance, sociometric
analysis, etc. It is not strange then that this cluster is the closest
one to the cluster of physics papers. At this point, both S and DB
diverges in the selection of the other clusters. Silhouette identifies
two more clusters, one formed by 51 nodes (violet nodes in Fig.\ref{Small_World})
and another of 29 nodes (green nodes in Fig.\ref{Small_World}).
However, DB identifies only one cluster that consists of the union
of the two previously described clusters. It should be noticed that
the difference in the Silhouette parameter for the partition into
3 and 4 clusters is marginal, i.e., $S=0.8574$ for 4 clusters vs.
$S=0.8569$ for 3 ones. However, as we want to know the finest-grain
structure of this network we adopt here the division into 4 clusters.
The cluster formed by 51 nodes is mainly formed by papers about organizational
theory and the structure of social structures. The smallest cluster
is mainly about papers on areas of applications, including medicine
and epidemiology, psychology, information sciences, economics and education.
Then, it is not rare that this cluster appears a little bit more isolated
from the physics as well as from the mainstream sociology ones.

This example teaches us a fundamental difference between the density-based
methods and the method presented here, which is based on communicability
angles. For instance, the cluster 2 which is formed by 45 nodes has
50 edges inside the cluster. However, it has 78 edges connecting this
cluster with cluster 3 and 63 edges connecting it with cluster 1.
More critical is the case of cluster 4, the one having 29 nodes. It
has 33 edges between nodes inside the cluster. However, it has 99
edges connecting this cluster with cluster 3. This makes that density-based
methods split the network into a larger number of clusters. For instance,
Infomap identifies 11 clusters in this network, from which 3 clusters
are formed by only two nodes each and one cluster is formed by 3 nodes
only. In contrast, the average communicability angle between pairs
of nodes inside each of the 4 clusters identified here are significantly
smaller than those between the nodes in different clusters (see Table
\ref{inter_angles}). 

\begin{table}[]
\begin{centering}
\begin{tabular}{|c|c|c|c|c|}
\hline 
 & $C_{1}$ & $C_{2}$ & $C_{3}$ & $C_{4}$\tabularnewline
\hline 
\hline 
$C_{1}$ & 0.992 & 4.601 & 7.596 & 12.491\tabularnewline
\hline 
$C_{2}$ &  & 1.066 & 3.019 & 7.919\tabularnewline
\hline 
$C_{3}$ &  &  & 0.968 & 4.930\tabularnewline
\hline 
$C_{4}$ &  &  &  & 1.208\tabularnewline
\hline 
\end{tabular}
\par\end{centering}
\caption{Average inter- and intra-cluster communicability angle for the clusters obtained using Silhouette method in the "Small-World" citation network.}
\label{inter_angles}
\end{table}

\subsubsection{Analysis of a gene-gene network}

Now, we move to the analysis of a genetic network in which nodes represent genes that have been
identified in relation to a human disease \cite{human_diseases}.
Two nodes are connected if the corresponding genes are involved in
at least one common disease. In total there are 22 human diseases
studied and a class of genes involved in mixed diseases, which was named in \cite{human_diseases} as the "grey" class. In
this case we have a previous information about certain ``clusters''\textendash not
necessarily topological in their nature\textendash that exist in the
network. We also know the finest grain structure of those clusters
as we can know the specific disease in which every gene is involve
in. For instance, one of the 22 disease categories consists of genes
involved in cancer. Additionally, we could know which types of cancer
compose that category creating a much larger number of categories
or clusters for that network. Then, our goal here is not to study
the finest-grain structure of that network but to find those clusters
that group together some of these categories, possibly indicating interrelation
between genes in several diseases. That is, we are interested here
in finding a limited number of clusters in relation to the number
of disease classes considered. In practical terms we will limit here
the number of clusters to a maximum which is smaller than the number
of groups existing, i.e., 22.

\begin{figure}[]
\centering
\subfloat[]{\includegraphics[width=0.32\textwidth]{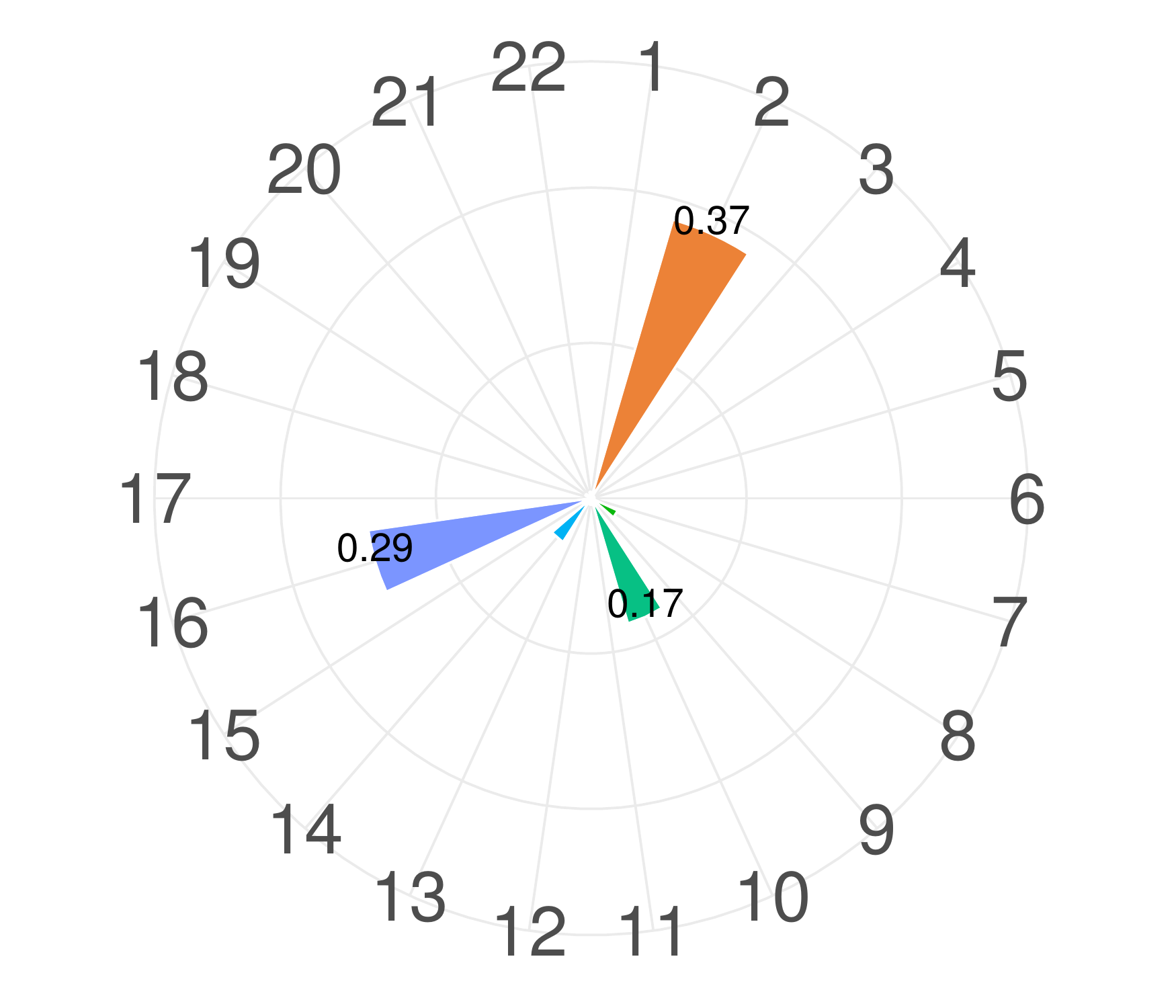}}
\subfloat[]{\includegraphics[width=0.32\textwidth]{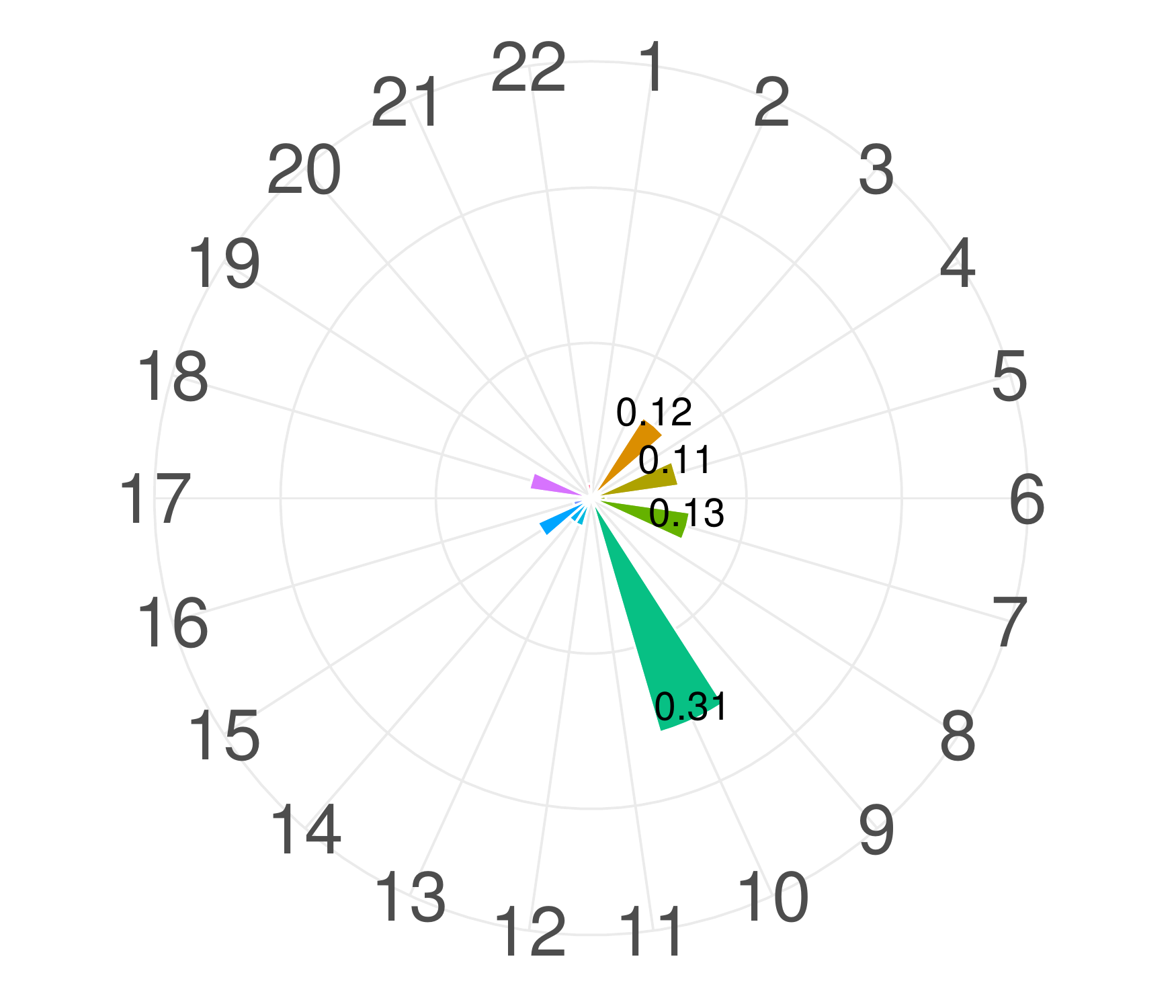}}
\subfloat[]{\includegraphics[width=0.32\textwidth]{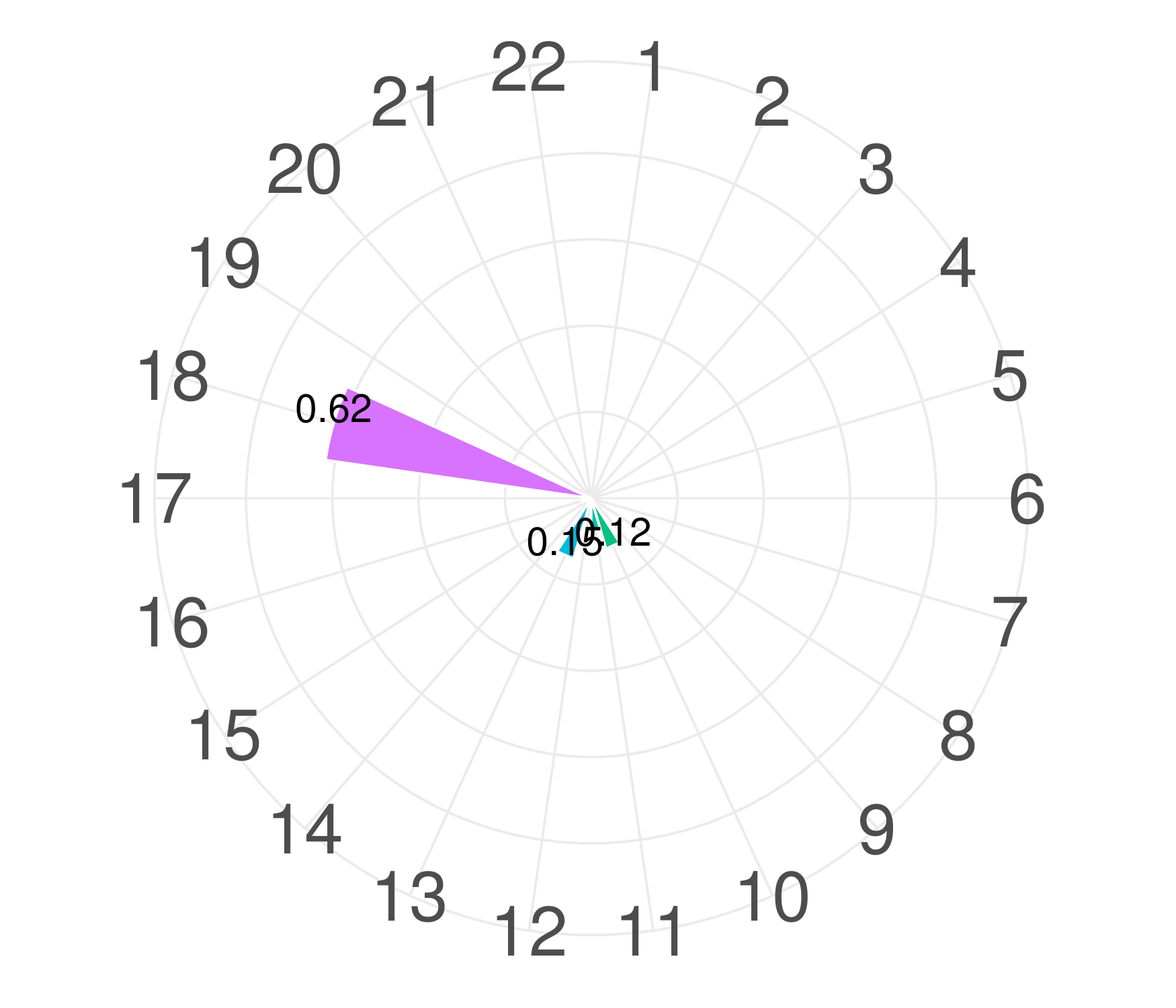}}
\hspace{10mm}
\subfloat[]{\includegraphics[width=0.32\textwidth]{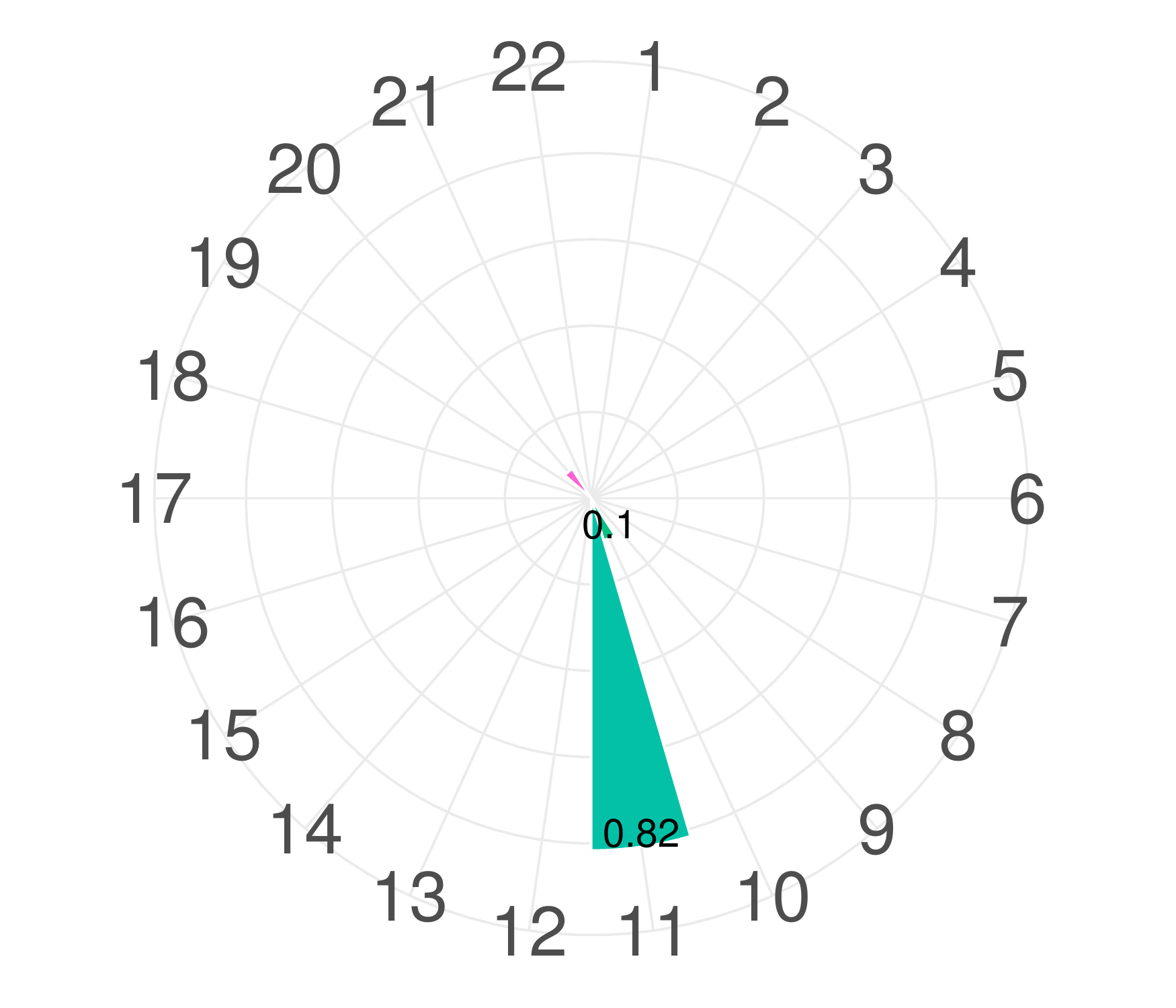}}
\subfloat[]{\includegraphics[width=0.32\textwidth]{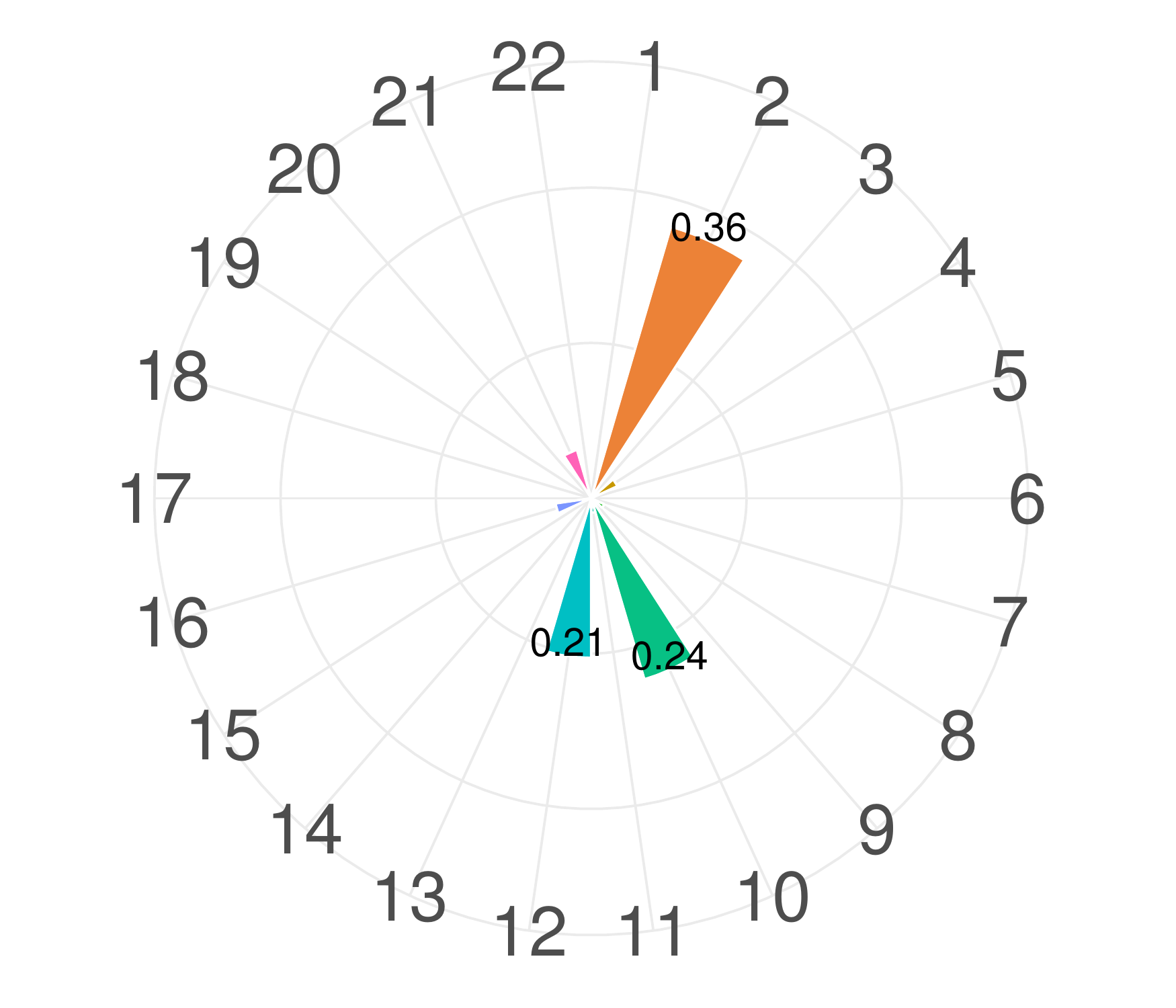}}
\subfloat[]{\includegraphics[width=0.32\textwidth]{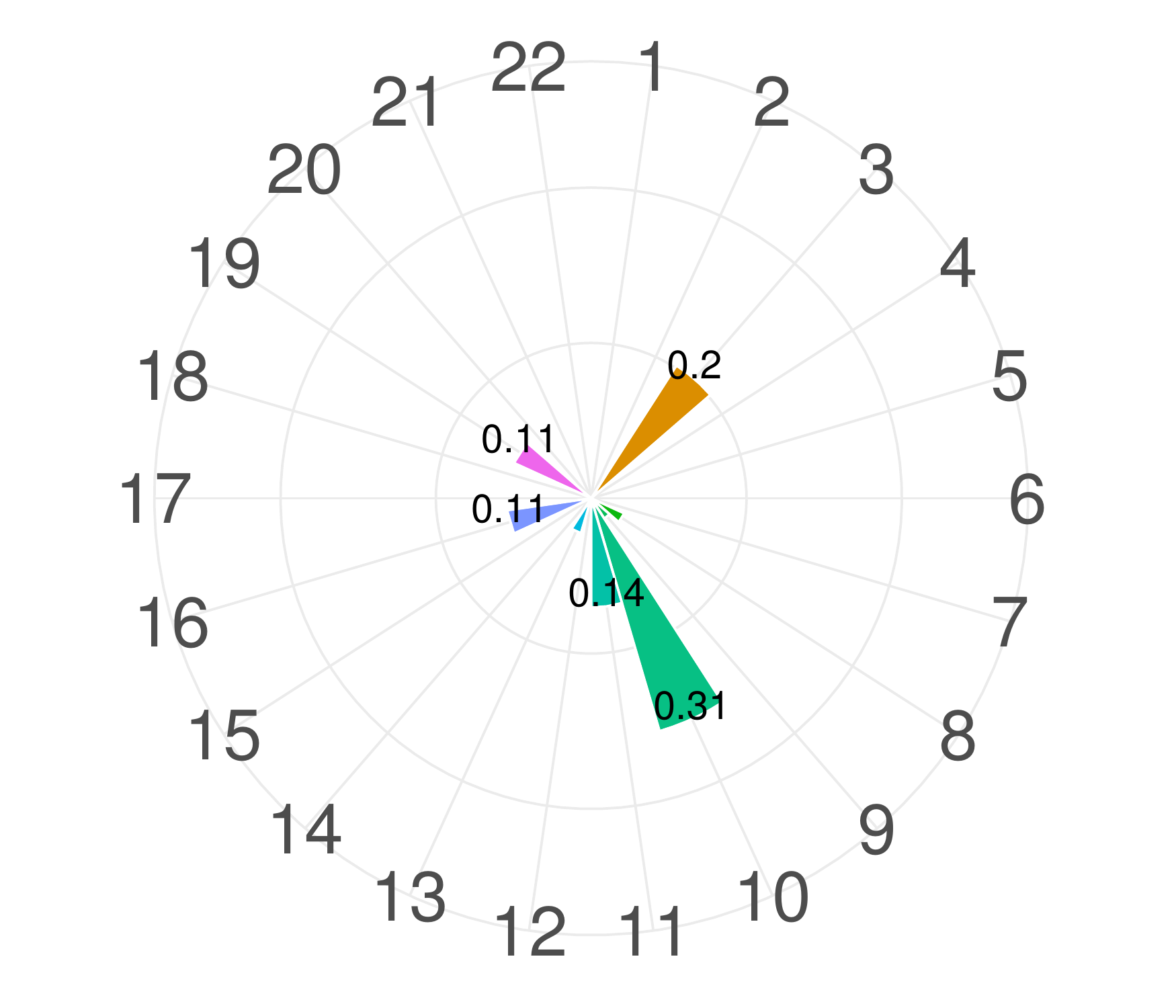}}
\hspace{10mm}
\subfloat[]{\includegraphics[width=0.32\textwidth]{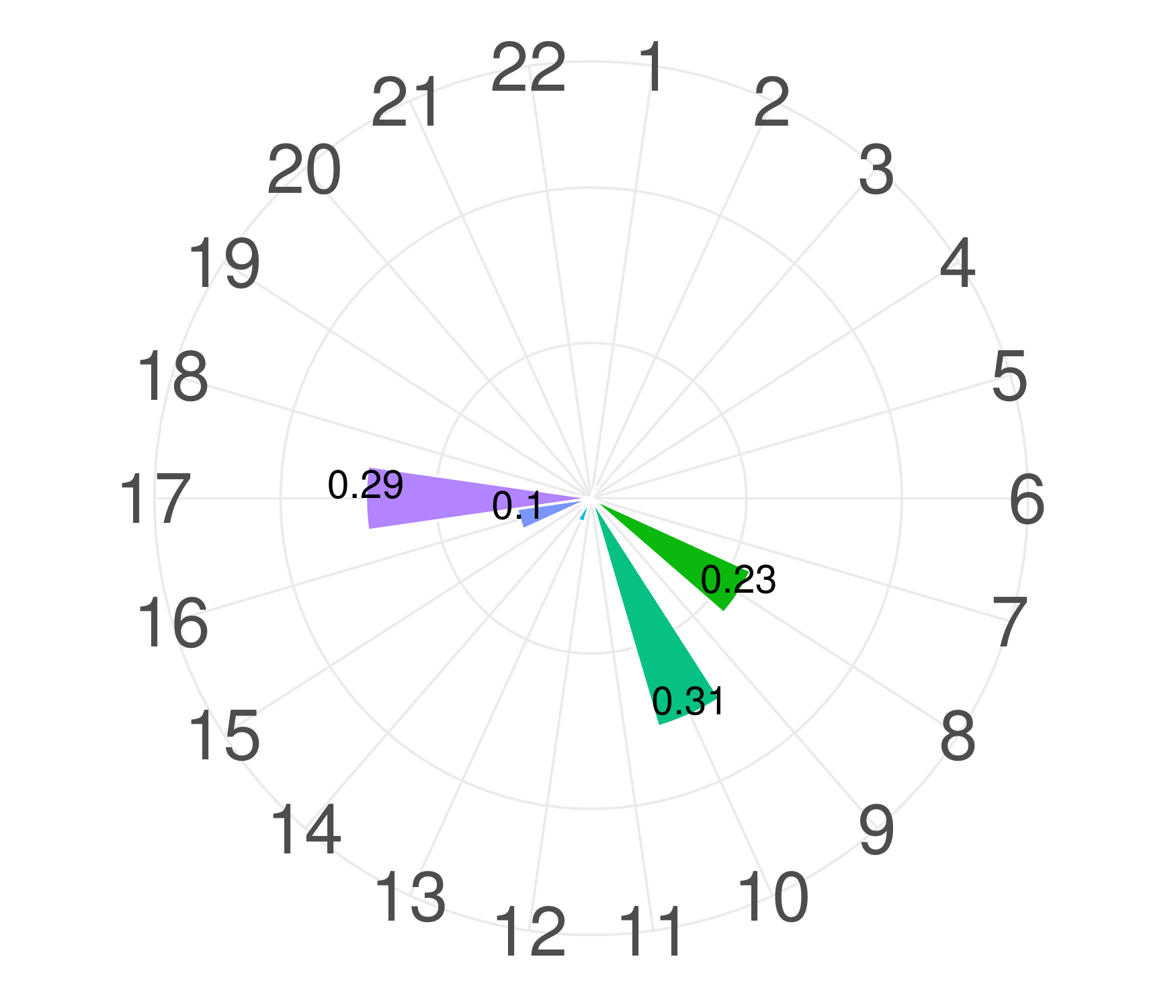}}
\subfloat[]{\includegraphics[width=0.32\textwidth]{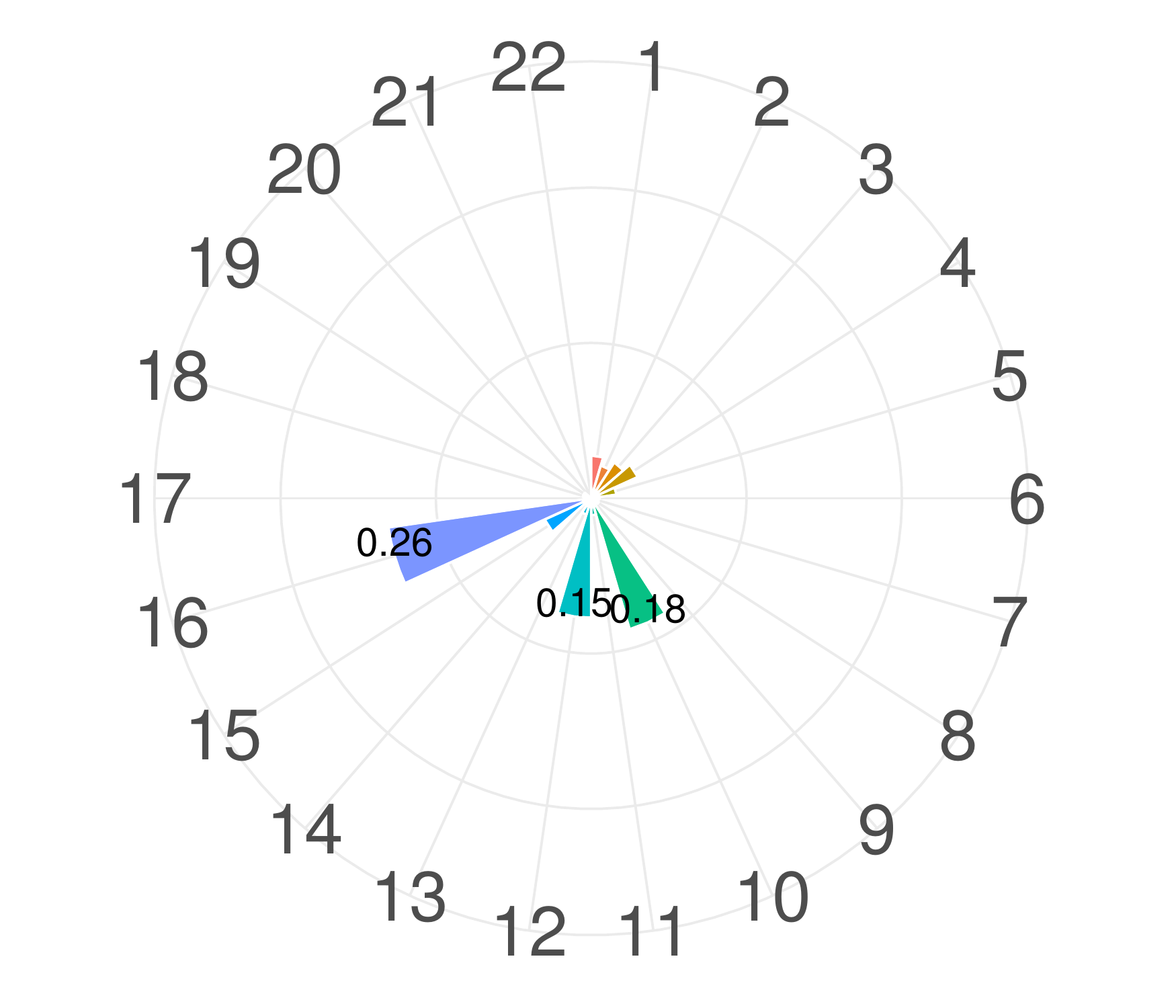}}
\caption{Radial histogram of the contribution of each of the 22 human diseases
considered to the clusters of genes obtained in this work. The diseases
are represented by a number according to the following code: 1: Bone;
2: Cancer; 3: Cardiovascular; 4: Connective tissue; 5: Dermatological;
6: Developmental; 7: Ear/Nose/Throat; 8: Endocrine; 9: Gastrointestinal;
10: Mixed; 11: Hematological; 12: Immunological; 13: Metabolic;
14: Multiple; 15: Muscular; 16: Neurological; 17: Nutritional; 18:
Ophtalmological; 19: Psychiatrical; 20: Renal; 21: Respiratory; 22:
Skeletal.}
\label{clusters genes}
\end{figure}

In this case Silhouette identifies 7 clusters and Davies-Bouldin identifies
8 ones (CH fails again by similar reasons as the ones described before). Some of the clusters identified by both CVIs are very similar
to each other. Indeed, the Fowlkes\textendash Mallows index \cite{Fowlkes} between
the two clustering gives a value of 0.9772 which indicates 97\% of
similarity between the two clustering. Then, deciding which of the
two clustering is the ``best'' is not a vital quest\textcolor{black}{ion
here. In selecting one of the two clustering we are inclined to the
one produced by the Davies-Bouldin Index, because it evaluates intra-cluster
similarity and inter-cluster differences instead of the Silhouette
Index which measures the distance between each data point, the centroid
of the cluster to which it was assigned to and the closest centroid
belonging to another cluster.}\textcolor{red}{{} }\textcolor{black}{We
then proceed to the analysis of this clustering of the genes involved
in human diseases. The first observation is that there are two clusters
mainly related to cancer diseases, one (Fig.\ref{clusters genes}
(a)) containing 37\% of the genes in the cluster related to cancer
and 29\% of genes related to neurological diseases and the other (Fig.\ref{clusters genes} (e)) with 36\% of genes involved in cancer,
21\% in immunological diseases and 24\% of mixed involvement. Cluster
2} \textcolor{black}{(Fig.\ref{clusters genes} (b))} has almost
equal contributions from genes involved in Cardiovascular, Dermatological
and Ear/Nose/Throat diseases and 31\% of genes with mixed involvement.
Cluster3 \textcolor{black}{(Fig.\ref{clusters genes} (c))} is mainly
formed by genes involved in ophtalmological diseases and cluster 4
\textcolor{black}{(Fig.\ref{clusters genes} (d)) is mainly populated
by genes related to }hematological disorders. Clusters 6-8 have participation
of genes involved in cardiovascular, neurological and psychiatric
diseases, nutritional and endocrine disorders as well as neurological
and immunological ones. 

We are not going to make an exhaustive analysis of each of the clusters
previously found but we will conduct a couple of experiments in order
to indirectly validate the biological significance of the clusters
found. The dataset used for the creation of the gene-gene network
compiled here used information about the involvement of genes in human
diseases as reported until 2007, in which the paper of Goh et al. \cite{human_diseases}
was published. Then, we selected two of the clusters found here for
our further analysis. The first is cluster 1 which is formed by 239
genes, two thirds of which are involved either in cancer or in neurological
diseases. According to the criterion used by Goh et al. \cite{human_diseases} in building
their network, a gene classified in the disease class ``neurological''
is not involved (as reported until 2007) in any other non-neurological
disorder. Otherwise, it is grouped as ``grey''. This means that
a gene grouped in the class of ``neurological diseases was not known
in 2007 to be involved also in ``cancer''. However, our clustering
method groups together 88 genes involved in cancer with 69 genes involved
in neurological disorders in the same cluster. Consequently, we formulate
the hypothesis that:

\textbf{Hypothesis 1. }Genes involved in neurological disorders which
are in cluster 1 can also be involved in cancer due to certain similarity
in their topological environment in the gene-gene network.

In order to test this hypothesis we carried out a bibliographic search
for all the genes in cluster 1 which were involved in neurological disorders
to find whether they have been recently reported in cancer. Our search
is based on the scientific literature published since 2006, a year
before the year in which Goh et al. \cite{human_diseases} paper was
published. In Table \ref{neuro_cancer} we report 19 genes that are
known to be related to neurological diseases such as ataxia, Amyotrophic
Lateral Sclerosis (ALS), Parkinson, epilepsy and Charcot-Marie-Tooth
disease, which are in cluster 1. In addition, in the most recent literature we have found
evidences of the involvement of these genes in breast, colorectal,
ovarian, lung, prostate, and other types of cancer. Although our search
has not being exhaustive, the current findings point out to the fact
that the clusters found in this work may contain important biological
information. In particular, that genes clustered in the same group
but having involvement in different diseases may have some ``promiscuity''
in the sense of being responsible for several diseases, such as the
ones reported in Table \ref{neuro_cancer} for neurological diseases
and cancer (all the references for the new findings are reported in
a Supplementary Information accompanying this paper). We should point
out that there are other 50 genes in this cluster which were reported
as involved in neurological diseases and for which we did not find any
connection with cancer. These genes could be target of experimental
search for involvement in different types of cancer also confirming
our hypothesis 1.

\begin{table}[h]
\begin{centering}
\resizebox{0.7\textwidth}{!}{

\begin{tabular}{|c|c|c|c|}
\hline 
No. & gene & neurological & cancer\tabularnewline
\hline 
\hline 
1 & NDRG1 & Charcot-Marie-Tooth & breast\tabularnewline
\hline 
2 & FGF14 & Spinocereballar ataxia & breast\tabularnewline
\hline 
3 & NEFH & ALS & breast\tabularnewline
\hline 
4 & PPP2R2B & Spinocereballar ataxia & breast\tabularnewline
\hline 
5 & SLC25A22 & epilepsy & breast, colorectal, \tabularnewline
\hline 
6 & GABRA1 & epilepsy & colorectal\tabularnewline
\hline 
7 & JPH3 & Huntington's & colorectal\tabularnewline
\hline 
8 & GJB1 & Charcot-Marie-Tooth & colorectal\tabularnewline
\hline 
9 & UCHL1 & Parkinson & colorectal, ovarian\tabularnewline
\hline 
10 & DNM2 & Charcot-Marie-Tooth & ovarian\tabularnewline
\hline 
11 & TDP1 & Spinocereballar ataxia & lung\tabularnewline
\hline 
12 & SOD1 & ALS & lung\tabularnewline
\hline 
13 & PARK7 & Parkinson & lung, prostate\tabularnewline
\hline 
14 & LRRK2 & Parkinson & non-skin\tabularnewline
\hline 
15 & KIF1B & Charcot-Marie-Tooth & hepatocellular carcinoma\tabularnewline
\hline 
16 & HSPD1 & Spastic ataxia/paraplegia & colon\tabularnewline
\hline 
17 & NR4A2 & Parkinson & gastroinstestinal\tabularnewline
\hline 
18 & Rab7 & Charcot-Marie-Tooth & thyroid adenoma\tabularnewline
\hline 
19 & SNCAIP & Parkinson & medulloblastome\tabularnewline
\hline 
\end{tabular}
}
\par\end{centering}
\caption{List of genes which were grouped in cluster 1 due to their involvement
in neurological diseases and which have been recently reported to be
involve in cancer as correspond to 37\% of genes in cluster 1. }

\label{neuro_cancer}
\end{table}

The second example is related to cluster 5, which is formed by 107
genes from which 1/3 of genes are involved in cancer and a quarter
of genes were classified in the group ``grey''. The group grey was
created by Goh et al. \cite{human_diseases} for grouping all those
genes which were involved in more than one type of human disease group.
That is, if a gene is involved in two different kinds of neurological
diseases, e.g. ataxia and paraplegia, it is still considered as in
the class of neurological diseases. However, a gene like APOA1, which
is involved in Alzheimer disease (neurological), Hyperlipoproteinemia
(metabolic), and Myocardial infarction (cardiovascular), is
considered in the class ``grey''. In cluster 5 there are 26 genes
in class ``grey''. From these 26 genes, 12 genes were already reported
to be involved in cancer by Goh et al. \cite{human_diseases} when
they grouped them in the class ``grey''. Then, there are 14 genes
in this cluster which are involved in different diseases but that
were not reported to be involved in cancer by Goh et al. \cite{human_diseases}.
Consequently, we elaborate our second hypothesis.

\textbf{Hypothesis 2. }Genes involved in multiple diseases which are
in cluster 5 can also be involved in cancer due to certain similarity
in their topological environment in the gene-gene network.

In Table \ref{cancer} we report our findings for 10 out of the 14
remaining ``grey'' genes in cluster 5, which have been recently
reported to be involved in several types of cancer. All the references
for the new reports are given in a Supplementary Information accompanying
this paper. Here again these results indirectly validate the biological
significance of the clusters found in this work using unsupervised
learning based on networks embedded into hyperspheres. 

\begin{table}[h]
\begin{centering}
\resizebox{\textwidth}{!}{
\begin{tabular}{|c|c|>{\centering}p{8cm}|c|}
\hline 
No. & gene & ``grey'' diseases & cancer\tabularnewline
\hline 
\hline 
1 & ABCA1 & Cerebral amyloid angiopathy, Coronary artery disease, HDL cholesterol
level QTL, Tangier disease & prostate cancer\tabularnewline
\hline 
2 & ESR1 & Estrogen\_resistance, HDL cholesterol level QTL, Migraine & hormone-resistant metastatic breast cancer\tabularnewline
\hline 
3 & ALOX5 & Asthma, Atherosclerosis & chronic myeloid leukemia\tabularnewline
\hline 
4 & IL10 & Graft -versus-host disease, HIV, Rheumatoid arthritis & prostate cancer\tabularnewline
\hline 
5 & IL13 & Allergic rhinitis, Asthma & colon cancer\tabularnewline
\hline 
6 & CIITA & Bare lymphocyte syndrome, Multiple\_sclerosis, Rheumatoid arthritis & lymphoid cancers\tabularnewline
\hline 
7 & PTPRC & Multiple sclerosis, Severe combined immunodeficiency & lymphoblastic leukemia\tabularnewline
\hline 
8 & BDNF & Central hypoventilation\_ syndrome, Memory impairment, Obsessive-compulsive
disorder & lung cancer\tabularnewline
\hline 
9 & PLA2G7 & Asthma, Atopy, Platelet defect/deficiency & prostate cancer\tabularnewline
\hline 
10 & CD36 & Malaria, Platelet defect/deficiency & glioblastoma\tabularnewline
\hline 
\end{tabular}
}
\par\end{centering}
\caption{List of genes grouped in cluster 5 which were reported as being involved
in several diseases (``grey'' diseases) but not in cancer until
2007. All these genes have been recently found as being involved in
different types of cancer in correspondence with the majority of genes
in this cluster.}

\label{cancer}
\end{table}

\section{Conclusions}
In this work we propose a way to extract network information by considering
machine learning techniques applied over a Euclidean hyperdimensional
representation of relational data. It should be remarked that this
``geometric learning'' approach differs from others in the literature
in the following. While many geometric learning methods are based
on imposed embeddings of the network in given spaces, here we exploit
a natural embedding of the graph emerging from the flow of items through
its nodes and edges. This space is a Euclidean $(n-1)-$sphere, where
$n$ is the number of nodes of the graph. Here we have used geometric
unsupervised learning approaches to find clusters in networks. In
these clusters the nodes are grouped not by their intra- and inter-cluster
densities, but on the basis of their capacity of successfully delivering
items through the network. We also used nonmetric multidimensional
scaling to reduce the dimensionality of the hyperspheres in which
the networks are naturally embedded to 3-dimensional ones that allow
visualization of the systems represented. It is important to remark
that the value of the current work is not in the sophistication of
the machine learning techniques used but in the novelty of this representation
systems of networks. Thus, further development using more advanced
machine learning and deep learning techniques will surely open new
possibilities for extracting high-quality structural information from
relational data, such as complex networks.



\bibliographystyle{siamplain}
\bibliography{refs_tags}

\begin{thebibliography}{10}

\bibitem{plo_blogs}
{\sc L.~A. Adamic and N.~Glance}, {\em The political blogosphere and the 2004
  {U}.{S}. election: divided they blog}, 2005.

\bibitem{Supervised}
{\sc F.~Azuaje}, {\em Witten {IH}, {F}rank {E}: {D}ata {M}ining: Practical
  machine learning tools and techniques 2nd edition}, BioMedical Engineering
  OnLine, 5 (2006), p.~51.

\bibitem{BA_model}
{\sc A.-L. Barab\'asi and R.~Albert}, {\em Emergence of scaling in random
  networks}, Science, 286 (1999), pp.~509--512.

\bibitem{Pajek_dataset}
{\sc V.~Batagelj and A.~Mrvar}, {\em Pajek datasets}, 2006.

\bibitem{networks_dynamics}
{\sc S.~Boccaletti, V.~Latora, Y.~Moreno, M.~Chavez, and D.~U. Hwang}, {\em
  Complex networks: Structure and dynamics}, Physics Reports, 424 (2006),
  pp.~175--308.

\bibitem{MDS}
{\sc I.~Borg and P.~Groenen}, {\em Modern Multidimensional Scaling: Theory and
  Applications}, Springer New York, 2013.

\bibitem{geometric_learning}
{\sc M.~M. Bronstein, J.~Bruna, Y.~LeCun, A.~Szlam, and P.~Vandergheynst}, {\em
  Geometric deep learning: Going beyond {E}uclidean data}, IEEE Signal
  Processing Magazine, 34 (2017), pp.~18--42.

\bibitem{Validation_CH}
{\sc T.~Caliński and J.~Harabasz}, {\em A dendrite method for cluster
  analysis}, Communications in Statistics, 3 (1974), pp.~1--27.

\bibitem{Validation_DB}
{\sc D.~L. Davies and D.~W. Bouldin}, {\em A cluster separation measure}, IEEE
  Transactions on Pattern Analysis and Machine Intelligence, PAMI-1 (1979),
  pp.~224--227.

\bibitem{Machine_learning_2}
{\sc P.~Domingos}, {\em A few useful things to know about machine learning},
  Commun. ACM, 55 (2012), pp.~78--87.

\bibitem{Grassman}
{\sc X.~Dong, P.~Frossard, P.~Vandergheynst, and N.~Nefedov}, {\em Clustering
  on multi-layer graphs via subspace analysis on grassmann manifolds}, IEEE
  Transactions on Signal Processing, 62 (2014), pp.~905--918.

\bibitem{Complex_networks_1}
{\sc E.~Estrada}, {\em The Structure of Complex Networks: Theory and
  Applications}, Oxford University Press, Inc., 2011.

\bibitem{Communicability}
{\sc E.~Estrada and N.~Hatano}, {\em Communicability in complex networks},
  Physical Review E, 77 (2008), p.~036111.

\bibitem{Angles}
{\sc E.~Estrada and N.~Hatano}, {\em Communicability angle and the spatial
  efficiency of networks}, SIAM Review, 58 (2016), pp.~692--715.

\bibitem{Estrada_Hatano_Benzi}
{\sc E.~Estrada, N.~Hatano, and M.~Benzi}, {\em The physics of communicability
  in complex networks}, Physics Reports, 514 (2012), pp.~89--119.

\bibitem{Estrada_Higham}
{\sc E.~Estrada and D.~J. Higham}, {\em Network properties revealed through
  matrix functions}, SIAM Review, 52 (2010), pp.~696--714.

\bibitem{Subgraph_Centrality}
{\sc E.~Estrada and J.~A. Rodr\'iguez-Vel\'azquez}, {\em Subgraph centrality in
  complex networks}, Physical Review E, 71 (2005), p.~056103.

\bibitem{Hyperspherical_embedding}
{\sc E.~Estrada, M.~G. S\'anchez-Lirola, and J.~A. D.~L. Pe\~na}, {\em
  Hyperspherical embedding of graphs and networks in communicability spaces},
  Discrete Appl. Math., 176 (2014), pp.~53--77.

\bibitem{Network_communities_1.5}
{\sc S.~Fortunato}, {\em Community detection in graphs}, Physics Reports, 486
  (2010), pp.~75--174.

\bibitem{Resolution_limit}
{\sc S.~Fortunato and M.~Barth\'elemy}, {\em Resolution limit in community
  detection}, Proceedings of the National Academy of Sciences, 104 (2007),
  pp.~36--41.

\bibitem{Network_communities_2}
{\sc S.~Fortunato and C.~Castellano}, {\em Community structure in graphs}, in
  Computational Complexity: Theory, Techniques, and Applications, R.~A. Meyers,
  ed., Springer New York, New York, NY, 2012, pp.~490--512.

\bibitem{Network_community_3}
{\sc S.~Fortunato and D.~Hric}, {\em Community detection in networks: A user
  guide}, Physics Reports, 659 (2016), pp.~1--44.

\bibitem{Fowlkes}
{\sc E.~B. Fowlkes and C.~L. Mallows}, {\em A method for comparing two
  hierarchical clusterings}, Journal of the American Statistical Association,
  78 (1983), pp.~553--569,
  \url{https://doi.org/10.1080/01621459.1983.10478008},
  \url{https://www.tandfonline.com/doi/abs/10.1080/01621459.1983.10478008}.

\bibitem{Gabriel}
{\sc K.~R. Gabriel and R.~R. Sokal}, {\em A new statistical approach to
  geographic variation analysis}, Systematic Biology, 18 (1969), pp.~259--278.

\bibitem{Unsupervised}
{\sc G.~Gan, C.~Ma, and J.~Wu}, {\em Data Clustering: Theory, Algorithms, and
  Applications}, Data Clustering: Theory, Algorithms, and Applications.

\bibitem{Girvan_Newman}
{\sc M.~Girvan and M.~E.~J. Newman}, {\em Community structure in social and
  biological networks}, Proceedings of the National Academy of Sciences, 99
  (2002), pp.~7821--7826.

\bibitem{human_diseases}
{\sc K.-I. Goh, M.~E. Cusick, D.~Valle, B.~Childs, M.~Vidal, and A.-L.
  Barabási}, {\em The human disease network}, Proceedings of the National
  Academy of Sciences, 104 (2007), pp.~8685--8690.

\bibitem{Clustering_validation_1}
{\sc M.~Halkidi, Y.~Batistakis, and M.~Vazirgiannis}, {\em On clustering
  validation techniques}, Journal of Intelligent Information Systems, 17
  (2001), pp.~107--145.

\bibitem{Clustering_3}
{\sc A.~K. Jain}, {\em Data clustering: 50 years beyond k-means}, Pattern
  Recognition Letters, 31 (2010), pp.~651--666.

\bibitem{Clustering_2}
{\sc A.~K. Jain and R.~C. Dubes}, {\em Algorithms for clustering data},
  Prentice-Hall, Inc., 1988.

\bibitem{Beta_2}
{\sc J.~W. Jaromczyk and G.~T. Toussaint}, {\em Relative neighborhood graphs
  and their relatives}, Proceedings of the IEEE, 80 (1992), pp.~1502--1517.

\bibitem{Machine_Learning_1}
{\sc M.~I. Jordan and T.~M. Mitchell}, {\em Machine learning: Trends,
  perspectives, and prospects}, Science, 349 (2015), pp.~255--260.

\bibitem{Clustering_1}
{\sc G.~Karypis, H.~Eui-Hong, and V.~Kumar}, {\em Chameleon: hierarchical
  clustering using dynamic modeling}, Computer, 32 (1999), pp.~68--75.

\bibitem{Clustering_4}
{\sc L.~Kaufman and P.~J. Rousseeuw}, {\em Finding Groups in Data: An
  Introduction to Cluster Analysis}, Wiley Series in Probability and
  Statistics, John Wiley \& Sons, 2005.

\bibitem{NMDS_2}
{\sc J.~B. Kruskal}, {\em Multidimensional scaling by optimizing goodness of
  fit to a nonmetric hypothesis}, Psychometrika, 29 (1964), pp.~1--27.

\bibitem{NMDS_1}
{\sc J.~B. Kruskal}, {\em Nonmetric multidimensional scaling: A numerical
  method}, Psychometrika, 29 (1964), pp.~115--129.

\bibitem{Complex_networks_2}
{\sc V.~Latora, V.~Nicosia, and G.~Russo}, {\em Complex Networks: Principles,
  Methods and Applications}, Cambridge University Press, 2017.

\bibitem{Outlier_1}
{\sc H.~Liu, S.~Shah, and W.~Jiang}, {\em On-line outlier detection and data
  cleaning}, Computers \& Chemical Engineering, 28 (2004), pp.~1635--1647.

\bibitem{Clustering_validation_2}
{\sc Y.~Liu, Z.~Li, H.~Xiong, X.~Gao, and J.~Wu}, {\em Understanding of
  internal clustering validation measures}, 2010.

\bibitem{Outlier_2}
{\sc C.-T. Lu, D.~Chen, and Y.~Kou}, {\em Algorithms for spatial outlier
  detection}, 2003.

\bibitem{dolphins}
{\sc D.~Lusseau, K.~Schneider, O.~J. Boisseau, P.~Haase, E.~Slooten, and S.~M.
  Dawson}, {\em The bottlenose dolphin community of {D}oubtful {S}ound features
  a large proportion of long-lasting associations}, Behavioral Ecology and
  Sociobiology, 54 (2003), pp.~396--405.

\bibitem{Hyperbolic}
{\sc L.~Ma, X.~Jiang, K.~Wu, Z.~Zhang, S.~Tang, and Z.~Zheng}, {\em Surveying
  network community structure in the hidden metric space}, Physica A:
  Statistical Mechanics and its Applications, 391 (2012), pp.~371--378.

\bibitem{Newman_review}
{\sc M.~E.~J. Newman}, {\em The structure and function of complex networks},
  SIAM Review, 45 (2003), pp.~167--256.

\bibitem{modularity}
{\sc M.~E.~J. Newman}, {\em Modularity and community structure in networks},
  Proceedings of the National Academy of Sciences, 103 (2006), pp.~8577--8582.

\bibitem{Spectral_clustering_2}
{\sc A.~Y. Ng, M.~I. Jordan, and Y.~Weiss}, {\em On spectral clustering:
  analysis and an algorithm}, 2001.

\bibitem{ER_model}
{\sc E.~Paul and R.~Alfréd}, {\em On random graphs {I}}, Publicationes
  Mathematicae (Debrecen), 6 (1959).

\bibitem{Association}
{\sc G.~Piatetsky-Shapiro}, {\em Discovery, analysis and presentation of strong
  rules}, in Knowledge Discovery in Databases, G.~Piatetsky-Shapiro and W.~J.
  Frawley, eds., AAAI Press, 1991, pp.~229--248.

\bibitem{Network_communities_1}
{\sc M.~A. Porter, J.~P. Onnela, and P.~J. Mucha}, {\em Communities in
  networks}, Notices of the American Mathematical Society, 56 (2009),
  pp.~1082--1097.

\bibitem{Validation_S}
{\sc P.~J. Rousseeuw}, {\em Silhouettes: A graphical aid to the interpretation
  and validation of cluster analysis}, Journal of Computational and Applied
  Mathematics, 20 (1987), pp.~53--65.

\bibitem{Graph_clustering}
{\sc S.~E. Schaeffer}, {\em Graph clustering}, Computer Science Review, 1
  (2007), pp.~27--64.

\bibitem{MachineLearningComplexNetworks}
{\sc T.~Silva and L.~Zhao}, {\em Machine Learning Complex Networks}, Springer
  International Publishing, 1~ed., 2016.

\bibitem{K-means}
{\sc D.~Steinley}, {\em K‐means clustering: A half ‐ century synthesis},
  British Journal of Mathematical and Statistical Psychology, 59 (2006),
  pp.~1--34.

\bibitem{NMI}
{\sc A.~Strehl and J.~Ghosh}, {\em Cluster ensembles --- a knowledge reuse
  framework for combining multiple partitions}, J. Mach. Learn. Res., 3 (2003),
  pp.~583--617.

\bibitem{Beta_skeleton}
{\sc G.~T. Toussaint}, {\em The relative neighbourhood graph of a finite planar
  set}, Pattern Recognition, 12 (1980), pp.~261--268.

\bibitem{Dimensionality_reduction_1}
{\sc L.~van~der Maaten, E.~Postma, and H.~Herik}, {\em Dimensionality
  Reduction: A Comparative Review}, vol.~10, 2007.

\bibitem{Spectral_clustering_1}
{\sc U.~von Luxburg}, {\em A tutorial on spectral clustering}, Statistics and
  Computing, 17 (2007), pp.~395--416.

\bibitem{Xiao_Hancock}
{\sc B.~Xiao and E.~R. Hancock}, {\em Geometric characterisation of graphs},
  Image Analysis and Processing – ICIAP 2005, Berlin, Heidelberg, 2005,
  Springer Berlin Heidelberg, pp.~471--478.

\bibitem{karate_club}
{\sc W.~Zachary}, {\em An information flow model for conflict and fission in
  small groups}, Journal of Anthropological Research, 33 (1977), pp.~452--473.

\bibitem{Machine_Learning_3}
{\sc L.~Zhou, S.~Pan, J.~Wang, and A.~V. Vasilakos}, {\em Machine learning on
  big data}, Neurocomput., 237 (2017), pp.~350--361.

\end{thebibliography}
\end{document}